\documentclass[pdflatex,sn-nature,nolineno]{sn-jnl}

\usepackage{graphicx}
\usepackage{amsmath,amssymb}
\usepackage{xcolor}

\unnumbered
\begin{document}

\title[]{Six objects reported at 480 mph in a declassified military infrared video show brightness flicker at bird wingbeat rates}

\author*[1]{\fnm{Jacob} \sur{Haqq-Misra}}\email{jacob@bmsis.org}
\author[2]{\fnm{Ravi} \sur{Kopparapu}}\email{ravikumar.kopparapu@nasa.gov}
\author[3]{\fnm{Richard} \sur{Cloete}}\email{richard.cloete@cfa.harvard.edu}
\author[3]{\fnm{Omer} \sur{Eldadi}}\email{omer.eldadi@cfa.harvard.edu}
\author[4]{\fnm{Michael} \sur{Shermer}}\email{mshermer@skeptic.com}
\author[3]{\fnm{Abraham} \sur{Loeb}}\email{aloeb@cfa.harvard.edu}

\affil*[1]{\orgname{Blue Marble Space}, \orgaddress{\city{Seattle}, \state{WA},
\country{USA}}}
\affil[2]{\orgname{NASA Goddard Space Flight Center}, \orgaddress{\city{Greenbelt},
\state{MD}, \country{USA}}}
\affil[3]{\orgname{Galileo Project, Harvard University, 60 Garden Street},
\orgaddress{\city{Cambridge}, \state{MA}, \country{USA}}}
\affil[4]{\orgname{The Skeptics Society}, \orgaddress{\city{Santa Barbara},
\state{CA}, \country{USA}}}

\abstract{In September 2026 the U.S. Department of War released an infrared video recorded from a military aircraft over the Middle East, with a mission report describing six small spherical objects moving at an estimated 480 miles per hour. The brightness of each object, measured in every frame, rises and falls steadily at 7.1--7.9 times per second, each at its own rate and out of step with the others. Planted constant-brightness objects show that neither compression nor the sensor produces this flicker. The same analysis of the Galileo Project's ground-based infrared video finds flying animals beating at 3.4--10.0 times per second, within the range of bird wingbeats. An island in the opening frames provides a scale for the scene. A slow object at about three-quarters of the aircraft's height would appear to cross the sea surface at the reported speed, and would be at most about 0.3 meters across. The six objects are best explained as a small flock of birds. The same analysis finds similar beats, at 2.3--11.8 times per second, in 27 objects in twelve other released military videos. Releasing the redacted sensor and platform data would provide information that could refute the bird hypothesis.}


\maketitle

\section{Introduction}\label{sec:intro}

The U.S.\ Department of War has been releasing records of unidentified anomalous phenomena (UAP) under its PURSUE program (Presidential Unsealing and Reporting System for UAP Encounters), which includes dozens of videos from infrared cameras on military aircraft~\cite{pursue}. Redactions to the on-screen displays restrict the information these released videos carry, which makes many of these videos unresolvable simply due to lack of data. None of the sensor videos released under PURSUE provide the range from the sensor to the object, which limits how well the path of an object in a video can constrain its velocity~\cite{haqqmisra2026limits}. Without knowing the range, a small object near the aircraft (such as a bird or balloon) can look the same as a large object farther away. Likewise, the aircraft's own motion can make a slow object appear fast. For example, in the ``Go Fast'' video recorded by the U.S.\ Navy in 2015, an object that appeared to skim the sea at high speed was shown to be far above the sea and moving at about the speed of the wind, based on the pointing angles and range displayed by the sensor in unredacted video footage~\cite{aaro}. However, the data that were available to resolve the Go Fast case are redacted in the PURSUE sensor videos.

We found another example of this effect in one of the PURSUE videos from the Department of War's sixth tranche of documents. DOW-UAP-PR135 was released on 18 September 2026 together with a mission report, DOW-UAP-D108, that had been filed by a U.S.\ Air Force crew for a sighting over the Middle East on 29 July 2025~\cite{pr135video,d108}. The report describes six small spherical objects, grouped together, that ``seemed to be agile, frequently changing direction'' and were ``able to change directions quickly and simultaneously.'' The report lists the estimated ``kinetic velocity'' as 480~mph (215~m/s) and the estimated altitude as flight level 220 (22{,}000~ft, or 6.7~km). The report also records no sensor engagement with the objects, which is consistent with the altitude value being an estimate rather than a measurement. In this analysis, we suggest that the six objects are a small flock of birds flying closer to the aircraft than assumed by the crew, with the apparent speed estimated at 480~mph most likely coming from the aircraft's own ground-relative motion. For observers familiar with infrared imagery, these six points that keep changing their arrangement may look like a small flock of birds; however, such an inference is not based on measurements. This study was not intended to either validate or refute the report. We consider the flock explanation as a hypothesis, and we test it with measurements that could have falsified this hypothesis.

Birds can be recognized in video by their wingbeat. A flapping wing changes the projected surface area of the bird facing the camera, which modulates the bird's brightness at the wingbeat frequency, or at twice that frequency. For birds in cruising flight, wingbeat frequencies range from about 2.5~Hz in the largest birds to more than 10~Hz in small songbirds~\cite{pennycuick1996,pennycuick2001,bruderer2010}. The frequency of such a beat does not depend on the range, so it can be measured even when the range is unknown. Wingbeats are an established way to recognize birds in radar echoes~\cite{zaugg2008,bruderer2010}. Infrared cameras have long been used to count birds that migrate at night~\cite{zehnder2001,gauthreaux2006,weisshaupt2016}. In infrared video the wingbeat has been proposed as a way to identify birds whose wings can be resolved~\cite{matzner2015}, and it has been measured from the changing outline of bats filmed at 131.5 frames per second~\cite{breslav2012}. The objects in DOW-UAP-PR135 are too small on the image to show their wings, so we measure the beat from their brightness alone.

Here we perform a frame-by-frame analysis of the brightness and apparent size of each of the six objects in DOW-UAP-PR135. Every object beats steadily at 7.1--7.9~Hz, each independent of the others. Analysis of each beat eliminates video compression, pixel sampling, and sensor defects as explanations. We then apply the same frame-by-frame analysis to infrared video of flying animals from the ground-based camera array of the Galileo Project near Las Vegas~\cite{loeb2023overview,domine2025} and find several examples of objects with beats at bird wingbeat frequencies. Finally, we use an island that the video holds in view for its first ten seconds to fix the scale and the viewing geometry. This gives the speed of the objects over the sea surface, the field of view, the height of the aircraft, and an upper limit on the size of the objects. A slow object at most about a quarter of a meter across, at about three-quarters of the aircraft's height, fits all of these. The same scene indicates a black-hot display, so the bright objects were colder than the sea, as birds at that height would be. Together, these results provide evidence that the objects in DOW-UAP-PR135 are a small flock of birds. Applied to the other sensor videos released under PURSUE, the same measurements find similar beats, at 2.3--11.8~Hz, in 27 objects across twelve videos. This analysis also underscores the value of beat-frequency analysis in the characterization of UAP, as the beat frequency does not depend on the range and can be applied even when sensor telemetry or aircraft altitude data are redacted.

\section{Results}\label{sec:results}

\subsection{\texorpdfstring{The video and the report}{The video and the report}}\label{sec:res_video}

\begin{figure}[t]
\centering
\includegraphics[width=\textwidth]{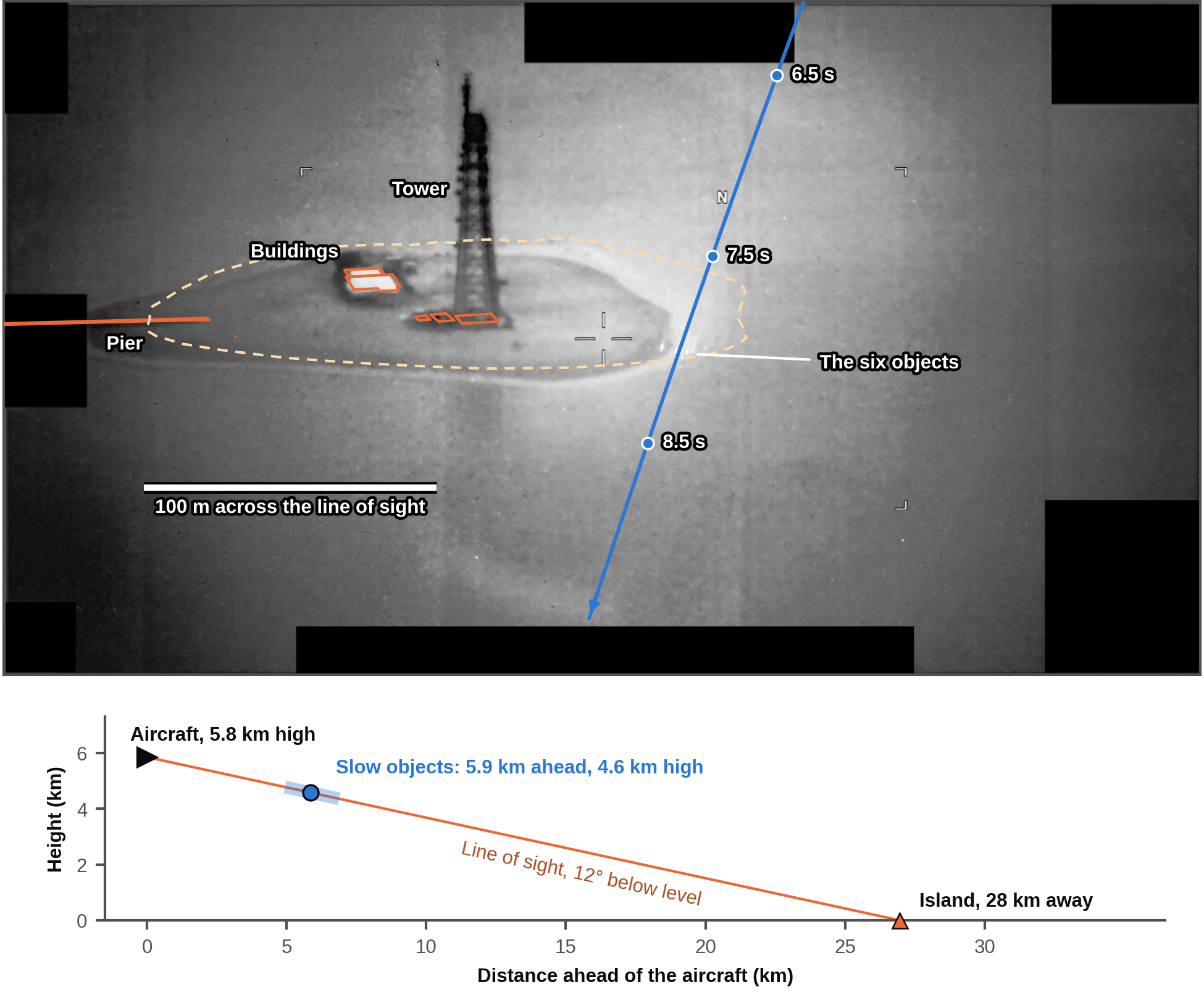}
\caption{Viewing geometry from the opening scene of DOW-UAP-PR135. Top: the frame at 8.0~s (1280 by 720 pixels; black boxes are redactions). Orange outlines are the mapped tower base, buildings, shelters, and pier of Qaruh Island~\cite{osm}, placed by the fitted camera. The dashed line is the mapped shore, which was not fitted. The bar is 100~m across the line of sight at the island. The blue line is the path of the objects across the scene from 6.2 to 9.3~s, with markers at 6.5, 7.5, and 8.5~s. Bottom: a side view, to scale, of the aircraft, its line of sight to the island, and the position of a slow object on that line. The blue bar shows the 68\% range.}
\label{fig:scene}
\end{figure}

DOW-UAP-PR135 is one minute of infrared video from the sensor turret of a military aircraft (Fig.~\ref{fig:scene}, top; see Data). Each frame is 1280 by 720 pixels, and black boxes cover the parts of the display that were redacted before release. For the first ten seconds the sensor holds a small island with a lattice tower, several buildings, and a pier. From 0:06 to 0:09 a tight group of bright points crosses this scene from the top of the frame to the bottom. The sensor then turns to follow the group over open water until 0:47, and the last seconds show a change of magnification with nothing in view. The accompanying report, DOW-UAP-D108, describes six small spherical objects and gives their speed as 480~mph, their direction as southeast, and their altitude as flight level 220, all marked as estimates. It gives the aircraft's track as northwest at 120 knots; the aircraft's altitude is redacted. The video can be viewed at its public source~\cite{pr135video} and in the archive of this paper (see Data availability).

\subsection{Beat frequencies of the six objects}\label{sec:res_beats}

\begin{figure}[t]
\centering
\includegraphics[width=\textwidth]{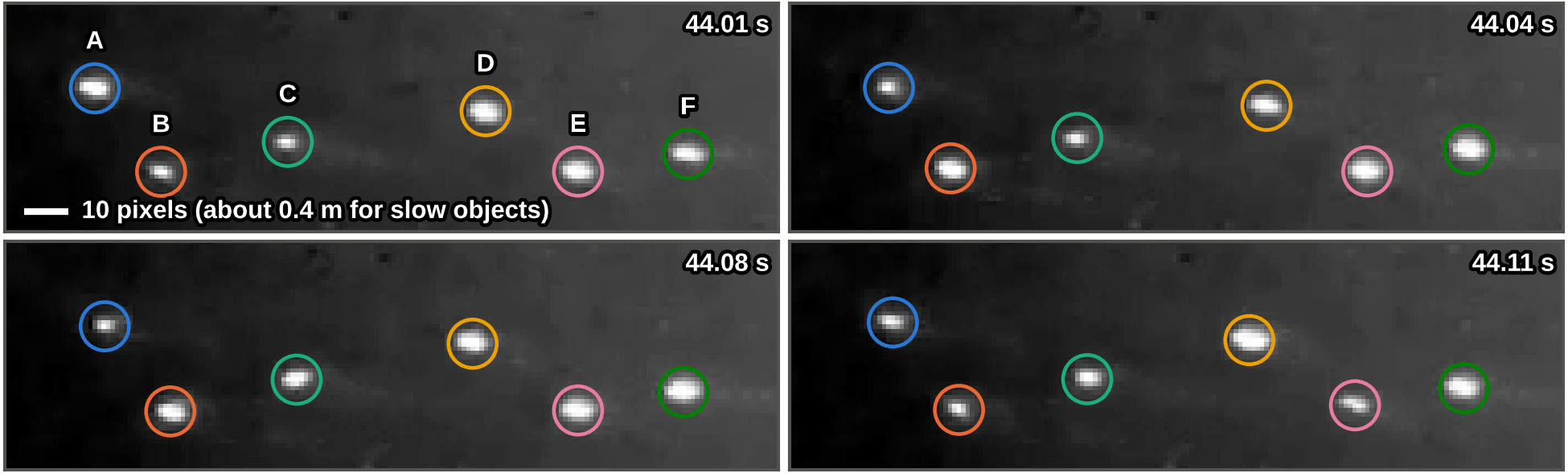}
\caption{Four consecutive frames of DOW-UAP-PR135 show the six objects (A--F) over about one beat (44.01--44.11~s). Each panel is a crop of 176 by 52 pixels from the 1280 by 720 pixel frame of the infrared video recorded from the aircraft. The bar is 10 pixels long. If the objects are slow, then the scale bar is about 0.4~m (see Ground-projected speed).}
\label{fig:frames}
\end{figure}

The objects move relative to one another by tens of pixels within 1--2~s, so that the group appears in turn as a line, a clump, pairs, and a loose ring. We tracked all six objects in every frame of the video from 41.3 to 46.3~s (Fig.~\ref{fig:frames}; see Tracking the six objects).

\begin{figure}[t]
\centering
\includegraphics[width=\textwidth]{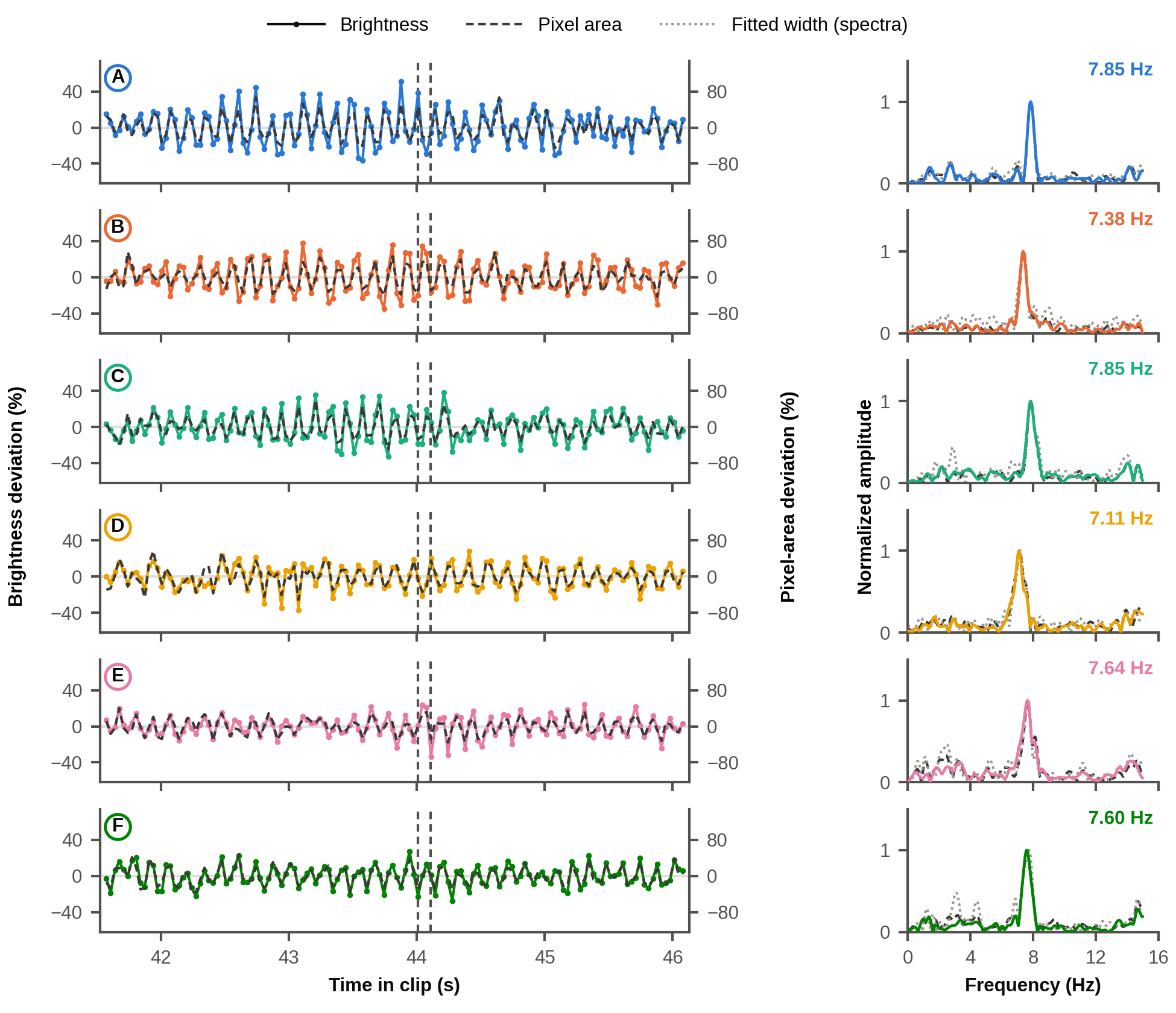}
\caption{Brightness and apparent size of the six objects beat at 7.1--7.9~Hz in the PURSUE video DOW-UAP-PR135. Left column: brightness (solid, left axis) and pixel area (dashed, right axis) of objects A--F, relative to 0.5~s running means; vertical lines mark the frames of Fig.~\ref{fig:frames}. Right column: normalized amplitude spectra of brightness (solid), pixel area (dashed), and fitted width (dotted).}
\label{fig:flicker}
\end{figure}

Each object brightens and dims steadily throughout this interval (Fig.~\ref{fig:flicker}). The brightness of the objects beats at 7.11--7.85~Hz, with amplitudes of 12--26\% about each running mean, and each beat stands 100--280 times above the median level of its spectrum. The same beat also appears in two measures of each object's size: its pixel area and its fitted Gaussian width. For every object, these three peak frequencies agree within 0.13~Hz (Fig.~\ref{fig:flicker}).

The six objects do not beat together, and their frequencies span 0.74~Hz. A and C beat at 7.85~Hz, E and F at 7.60--7.64~Hz, B at 7.38~Hz, and D at 7.11~Hz. Over the same frames, the brightness oscillations between any two objects are out of phase with each other. This phase difference between pairs ranges from 30$^\circ$ to 170$^\circ$ in magnitude, with eight pairs out of phase by more than 90$^\circ$. Even A and C, which show the same beat frequency, are 119$^\circ$ apart. This eliminates video flicker as an explanation, as a beat produced by the video itself would reach all six objects at one frequency and in one phase.

Dots of constant brightness, planted in the video at the objects' screen rate and encoded with the video's settings, show amplitudes of only 0.7--1.7\% in the 7.1--7.9~Hz band, against 12--26\% for the objects, and 0.6--0.9\% before encoding (see Controls). Neither the 7.49~Hz cadence of the video compression nor the measurement aperture therefore makes the beat. Ten fixed defects of the sensor, measured in the same way, have their strongest peaks at 1.5--3.3~Hz.

The core of each object is clipped, meaning that its pixels reach the maximum recorded value (see Apparent size). A brighter object can thus also appear larger, so a beat frequency in a measurement of size does not necessarily imply a change in size. This was tested by planting dots whose brightness beats but whose width is fixed; these show the same beat in pixel area, at 16--19\%. The size measures therefore confirm the beat frequency by independent measurements, but they do not show a change in the objects' size.

The position of each object also cycles at its own beat frequency. Measured against the mean position of the other five, which removes the motion of the camera, each object moves by 0.4--0.8 pixel at its beat frequency, and by no more than 0.12 pixel at half that frequency (see Position).

\subsection{Beat frequencies of birds in infrared video}\label{sec:galileo}

\begin{figure}[t]
\centering
\includegraphics[width=\textwidth]{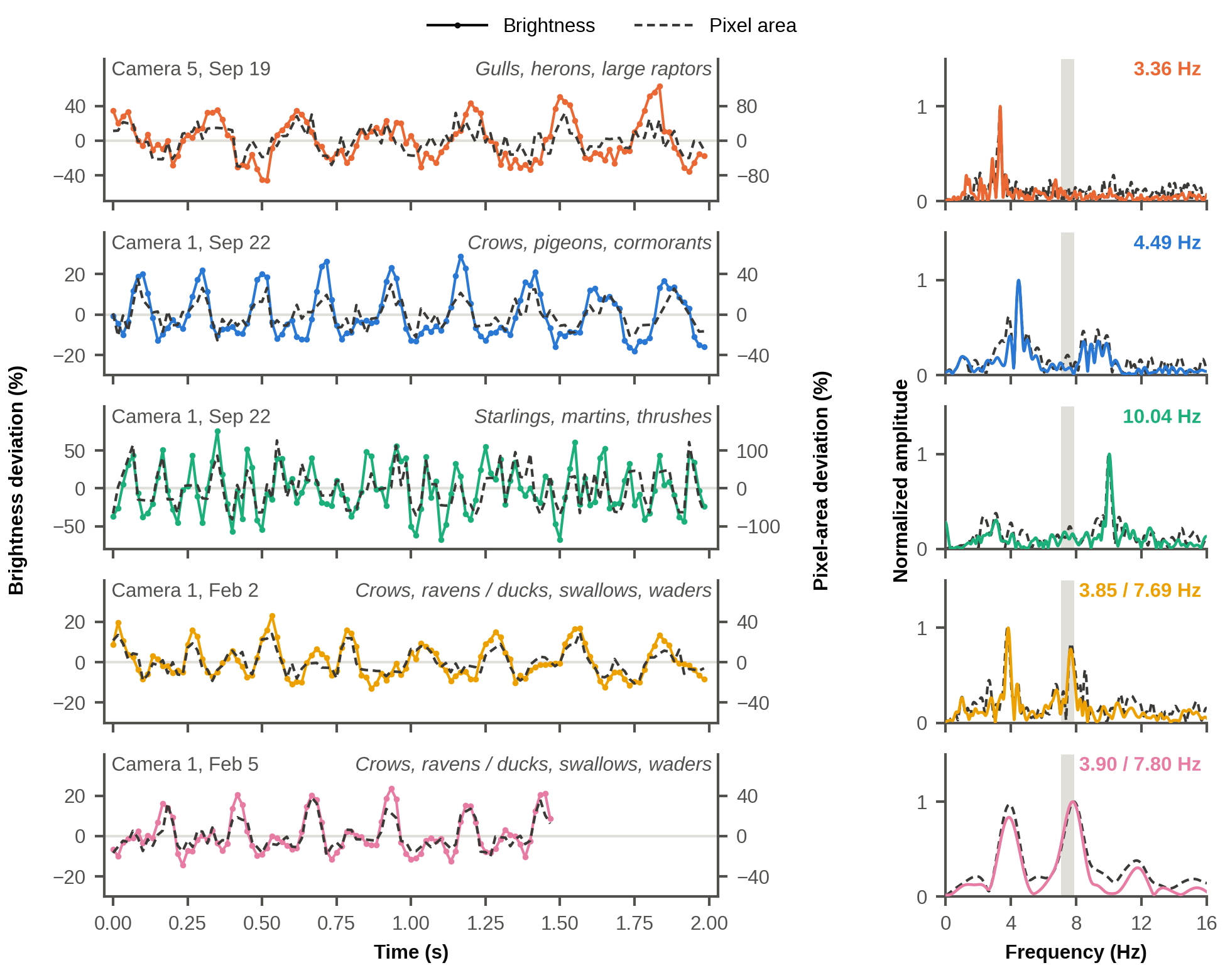}
\caption{Flying animals in infrared video beat in brightness and apparent size together in the Galileo Project data~\cite{loeb2026}. Left column: brightness (solid, left axis) and pixel area (dashed, right axis) of five flyers recorded by the Galileo Project, with the bird groups whose measured wingbeat frequencies match each flyer~\cite{bruderer2010}. Right column: normalized amplitude spectra of brightness (solid) and pixel area (dashed). The last two flyers brighten twice per wingbeat. The gray band indicates the frequency range of objects A--F.}
\label{fig:groups}
\end{figure}

A steady beat of this kind could be produced by flapping wings. To see whether flying animals produce a beat frequency in thermal video, and at what frequencies, we applied the same frame-by-frame analysis to infrared video of the sky recorded by the Galileo Project~\cite{loeb2026} (see Infrared video of flying animals). We readily found numerous flying animals on evenings and mornings, all of which beat steadily in brightness and size together during the entire time on camera. The rows of Fig.~\ref{fig:groups} show examples of these beats on a common time scale, alongside their spectra.

These five flyers beat at 3.4--10.0~Hz, which falls within the observed wingbeat frequencies of birds in cruising flight (from about 2.5~Hz to more than 10~Hz~\cite{pennycuick1996,pennycuick2001,bruderer2010}). The beat frequency can even provide constraints on the kinds of bird in a given observation, although this does not necessarily constrain particular species. Bats also beat their wings at about 3--14~Hz in free flight~\cite{norberg2012} and could explain some of these observations. This comparison shows that this frame-by-frame analysis method is able to recover a steady beat frequency from flying animals of size and brightness similar to the DOW-UAP-PR135 objects.

\subsection{\texorpdfstring{Scale and viewing geometry from the island}{Scale and viewing geometry from the island}}\label{sec:res_geometry}

The island provides a scale for the video. The tower, buildings, and pier match the mapped structures of Qaruh Island, Kuwait, seen from the southeast~\cite{osm} (Fig.~\ref{fig:scene}, top; see Viewing geometry). A camera fitted to these structures gives 0.32~m per pixel across the line of sight at the island (0.31--0.33~m; ranges here hold 68\% of the repeated fits), so the frame is about 410~m wide there. The line of sight points 12$^\circ$ below the horizontal (11--13$^\circ$), toward azimuth 325$^\circ$.

While the sensor views the island, the group crosses the scene in a straight line at 210 pixels per second, 19$^\circ$ to the left of straight down the screen. Projected onto the sea surface, this is 297~m/s, or 665~mph (626--705~mph), toward the southeast (azimuth 149$^\circ$). The crew's estimate of 480~mph toward the southeast is a speed of this magnitude, in the same direction.

The group's rate down the screen is not constant but accelerates by 6.5 pixels per second in each second (6.3--6.8). The approach of the aircraft produces such an apparent acceleration, and if the objects kept a steady velocity it fixes the angle that one pixel spans, whatever their range and speed (see Viewing geometry). The field of view is 0.85$^\circ$ across the frame (0.78--0.92$^\circ$). With it the island is 28~km from the aircraft (26--30~km), and the aircraft is 5.8~km above the sea (5.4--6.3~km), about 19{,}000~ft (Fig.~\ref{fig:scene}, bottom). The island's own growth on the screen, which needs no assumption about the objects, gives a field of view of 0.87$^\circ$ (0.72--1.09$^\circ$), in agreement (see Viewing geometry). This is an estimate, from the released pixels and a map, of the altitude that is redacted in the report.

\subsection{Parallax explains the reported speed}\label{sec:res_speed}

Birds do not fly at 480 mph. This suggests that the speed over the sea surface is an artifact of the parallax effect, or the apparent shift in an object's position relative to a distant background when viewed from different positions. The report marks the speed as estimated and does not record that a laser range was taken (see Data). Without a range measurement, the only range the targeting system can assign to a point in the image is the range to where the line of sight meets the ground. This means that a speed estimated from the imagery is the speed of that ground point, not of the object itself.

For an aircraft at height $h_{\rm air}$ and an object at height $h_{\rm obj}$ on the same line of sight, the ground-projected speed $v_{\rm gnd}$ gives the object's height for any speed $v_{\rm obj}$ that it is assumed to have (equation~\eqref{eq:vg}; see Ground-projected speed). Even a stationary object shows a ground-projected speed, directed opposite to the aircraft's track. This is the parallax effect: as the aircraft approaches, the line of sight through the object tilts more steeply, and its intersection with the ground moves back toward the aircraft.

The imagery is consistent with parallax as an explanation. While the sensor follows the group (0:11--0:47), the burned-in north pointer shows the line of sight turning by only 2.7$^\circ$, which implies that the aircraft was flying almost straight toward the objects. The group's track over the sea surface, toward the southeast, is opposite to the aircraft's track toward the northwest, and the report gives the same two directions. Because the sensor then follows the group, any turn of the sensor moves all six objects together on the screen, so changes of direction that appear simultaneous do not require the objects to turn together.

\begin{figure}[t]
\centering
\includegraphics[width=\textwidth]{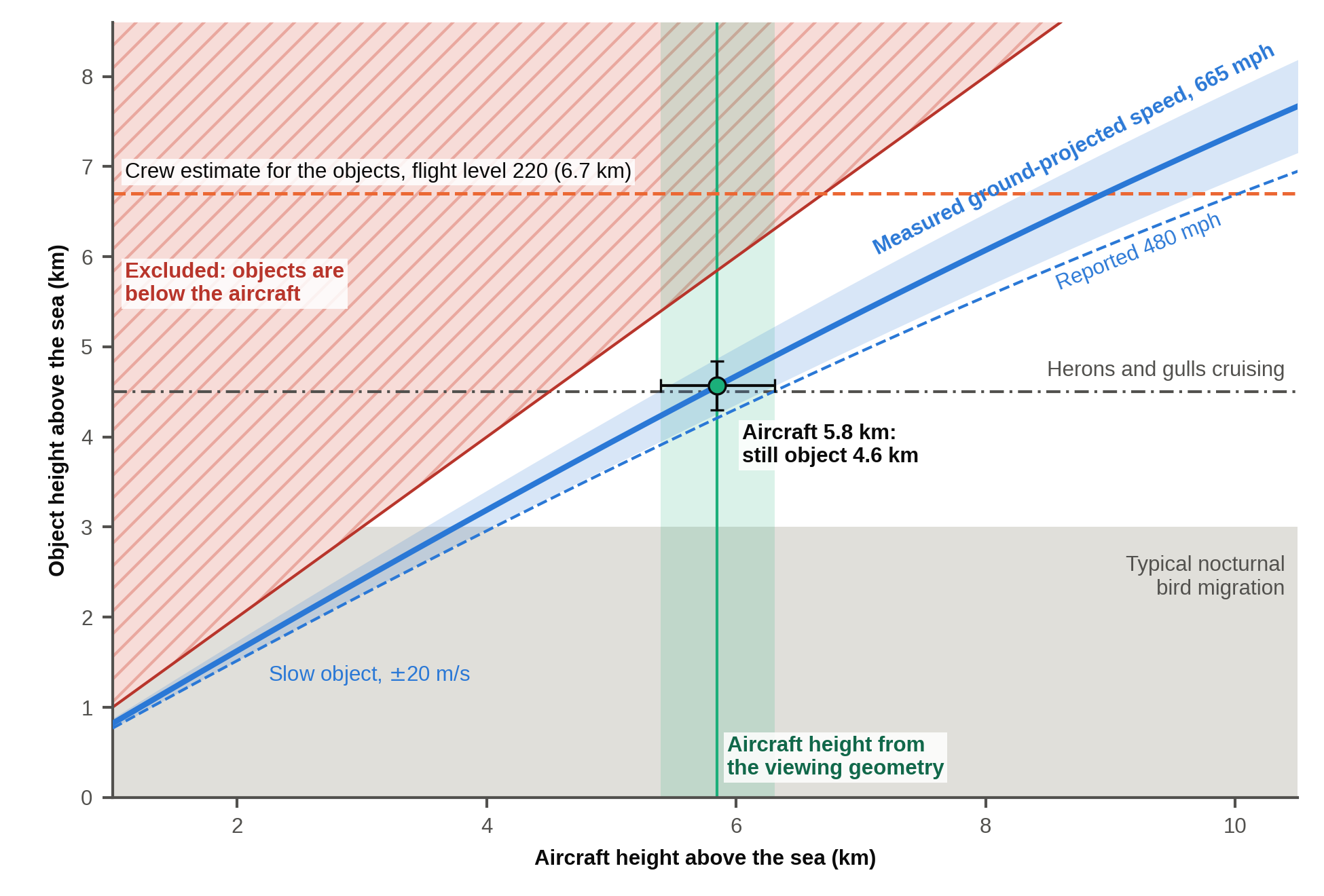}
\caption{A slow object near the aircraft reproduces the measured and the reported speed. Height of an object at which the ground-projected speed reads 665~mph, as measured against the island (solid blue line: stationary object; shaded band: object speeds of $\pm$20~m/s), or 480~mph, as reported (dashed blue line: stationary object), versus the aircraft's height above the sea (equation~\eqref{eq:vg}). The green band is the aircraft's height from the viewing geometry (68\% range), and the point marks a stationary object at that height. The hatched region, where the object would be above the aircraft, is excluded because the objects are seen against the sea. Altitudes are indicated for typical nocturnal bird migration~\cite{dokter2011} (gray band), herons and gulls cruising~\cite{schmaljohann2009} (dash-dot line), and the flight level 220 estimate provided by the crew (dashed orange line).}
\label{fig:height}
\end{figure}

Figure~\ref{fig:height} shows the solutions of equation~\eqref{eq:vg} for the measured ground-projected speed of 665~mph, and for the reported 480~mph. A stationary object ($v_{\rm obj} = 0$) shows the measured speed when it is at 78\% of the aircraft's height above the sea (76--81\%), and a slow object, moving at up to 20~m/s toward or away from the aircraft, at 73--83\%. With the aircraft at 5.8~km, a stationary object is at 4.6~km (4.3--4.8~km), 5.9~km ahead of the aircraft and 1.3~km below it (Fig.~\ref{fig:scene}, bottom). The reported 480~mph, in place of the measured speed, would put a stationary object at 72\% of the aircraft's height.

The same geometry bounds the objects' size. During the crossing each point is about as wide as the sensor's own blur (see Viewing geometry), and at the distance of a stationary object one pixel is 0.07~m. Each object is therefore at most 0.27~m across (0.24--0.30~m; 0.21--0.34~m for a slow object), neighbors are about 0.5~m apart, and the group is about 2~m wide. This bound applies to the width of each object across the frame, not to its true size. A bird shows its full wing span across the frame only when it flies toward or away from the camera; for a bird heading at an angle $\phi$ to the line of sight, a span $b$ appears as $b\,|\cos\phi|$. Because the heading is not known in this case, the bound limits the wing span only together with the angle. Of the 12 species that Bruderer et al.\ recorded beating their wings at 7.1--7.9~Hz (see Discussion), the swallows, with spans of 0.32--0.33~m, fit within 0.27~m at headings more than about 35$^\circ$ from the line of sight. The swifts (0.40~m) fit at more than about 48$^\circ$, the thrushes and the turtle dove (0.43--0.52~m) at more than 51--59$^\circ$, and the plover, godwit, nightjar, and duck (0.60--0.78~m) at more than 63--70$^\circ$~\cite{bruderer2010}. For a flock heading in a random direction these are chances of about 60\%, 47\%, 35--43\%, and 23--30\%. The bound also applies to the wings only if they, seen from 12$^\circ$ above, are as bright on the display as the body. As an alternative interpretation, an object at the sea surface, moving at the full 665~mph, could be up to 1.2~m across, in a group 9.3~m wide.

\section{Discussion}\label{sec:discussion}

Four results point to the same explanation. Each of the six objects beats steadily at 7.1--7.9~Hz, at its own frequency and phase, which neither the video nor the sensor produces. Flying animals recorded by infrared cameras beat in the same way, in brightness and apparent size together, at 3.4--10.0~Hz. The speed measured over the sea surface, and the speed and direction that the crew reported, are what a slow object at about three-quarters of the aircraft's height produces when no range is measured. And at that height each object is at most about 0.3~m across, in a group about 2~m wide. The simplest explanation is a small flock of birds, nearer to the aircraft than the ground point on which the speed was estimated. The number of objects, their changing arrangement, and the report's description of objects that were agile and changed direction are all consistent with this interpretation. The same measurements find similar beats, at 2.3--11.8~Hz, in 27 objects in twelve other videos released under PURSUE (Table~\ref{tab:beats}; see Other PURSUE videos), which suggests that birds may also account for some of the other objects in these releases.

The frequencies of the six objects lie within the range measured for birds in cruising flight~\cite{pennycuick1996,pennycuick2001,bruderer2010}. Of the 153 species measured by Bruderer et al.\ with tracking radar and cine camera, 12 have a median wingbeat frequency within 7.1--7.9~Hz in at least one flight condition. Among these are the northern shoveler, grey plover, bar-tailed godwit, turtle dove, common and pallid swifts, and several swallows and thrushes~\cite{bruderer2010}. Birds also beat their wings faster in thinner air. Among nocturnal migrants tracked by radar, the wingbeat frequency rose by 0.4~Hz per kilometer of altitude in flap-gliding birds and by 1.1~Hz per kilometer in bounding birds~\cite{schmaljohann2009}. This suggests that birds at 2--3~km could beat about 1--3~Hz faster than the same birds near sea level. The frequency alone therefore does not constrain a species. Six birds flying together would be expected to beat at similar but not identical rates and out of phase, as the six objects do.

The Galileo flyers also raise a question that the DOW-UAP-PR135 video alone cannot resolve. A bird's wing can brighten the image once per wingbeat or twice (on the downstroke and the upstroke). Two flyers, both recorded on February mornings, show the second case (last two rows of Fig.~\ref{fig:groups}). Their brightness rises twice in each cycle, strongly and then weakly, so their spectra have two
peaks of similar height, near 3.9~Hz and near 7.7--7.8~Hz (right-hand panels). Birds like these put strong brightness power within the 7.7--7.8~Hz frequency band while beating their wings at about 3.9~Hz; this is a rate typical of crows and ravens~\cite{bruderer2010}. In the DOW-UAP-PR135 objects, by contrast, the peak at half the beat frequency is only 10--17\% of the main peak. Their brightness therefore does not show the unequal double pulse of the February flyers, and their beat is most likely the wingbeat itself, at 7.1--7.9~Hz. However, a double pulse of two equal brightenings would leave no peak at half the frequency and cannot be excluded, in which case the wingbeat of the DOW-UAP-PR135 objects would be 3.6--4.0~Hz. Both options fall within the range of birds.

The position of each image offers one test that the present data allow. The image of a flapping bird shifts with its wings once per wingbeat. On the five Galileo flyers the position cycles most strongly at the wingbeat, including the flyer whose brightness is strongest at twice its wingbeat (see Position). The positions of the six objects cycle at their brightness beat frequencies and show no cycle at half those frequencies, which favors the direct reading, a wingbeat of 7.1--7.9~Hz. However, this is not decisive, and two examples of a double pulse, seen from below, are an insufficient calibration for six objects seen from in front and slightly above.

These two sources also differ in other ways. The Galileo cameras look up at their flyers and record long-wave infrared light, while the aircraft's sensor looked down at 12$^\circ$ and most likely recorded mid-wave infrared light (see Data). The brightness of a flapping bird beats because its projected area changes and because warmer parts of its body are exposed and covered. The balance between the two, and with it the number of brightenings per wingbeat, may differ with the viewing direction and the waveband. The Galileo flyers therefore show that flapping produces a steady beat at wingbeat rates in thermal video, but they do not show which harmonic a given view produces. The comparison in this study is based on the rate of the beat, which the wings establish in any waveband, and not on its depth.

Several further measurements could separate the two interpretations. The sensor's original frames, which are not clipped or contrast-enhanced, would preserve the shape of each brightness cycle: a double pulse would show as alternating stronger and weaker brightenings, or as power at half the beat frequency. A measured range would give the size of the objects and airspeed more accurately than the viewing geometry bounds used in this study. Birds that beat at 7--8~Hz, such as waders and ducks, have shorter wings and fly faster than the terns and gulls that beat at 3--5~Hz~\cite{bruderer2010}, and imagery that resolves the wings would show their position in each frame. And recordings of known species from above and from below, in both infrared bands, would show which birds and which views brighten once per wingbeat and which twice.

The location (see Viewing geometry) also supports this interpretation. The birds most likely to be flying near Qaruh Island are seabirds: five species of tern breed on Kuwait's offshore islands~\cite{edmonds2021}, and 5,000--6,000 terns nest on Kubbar, Umm al-Maradim, and Qaruh from May to July~\cite{seaalarm2018}, so the observation, on 29 July, falls at the end of the nesting season. Terns, gulls, and cormorants beat their wings more slowly
than the DOW-UAP-PR135 objects flicker. In cruising flight, the two terns measured by Bruderer et al.\ beat at 3.3 and 4.5~Hz, gulls at 2.7--3.8~Hz, and the great cormorant at 5.1~Hz~\cite{bruderer2010}. An increase of 0.4~Hz per kilometer, as in flap-gliding birds~\cite{schmaljohann2009}, would leave the faster tern below 6~Hz even at 3~km. A double pulse of two nearly equal brightenings per wingbeat, at 3.6--4.0~Hz, would therefore fit a tern. Birds that beat at 6.6--8.5~Hz, including ducks, grebes, and waders such as plovers and godwits~\cite{bruderer2010}, would fit the direct reading. Tracking radar in southern Israel recorded large numbers of nocturnal migrants with the continuous wingbeats of waders and waterfowl~\cite{bruderer1994}. The site thus allows both possibilities. The wing spans of these terns, waders, ducks, and grebes, 0.57--0.93~m~\cite{bruderer2010}, are more than twice the size bound of 0.27~m, so such birds fit it only if they flew within about 17--28$^\circ$ of across the line of sight, or if their wings did not show on the display (see Parallax explains the reported speed). Breeding common terns tracked by GPS stay at their colonies for nearly all of the night, although birds from one marine colony flew slowly over the sea after dark~\cite{kralj2026}, so a flock of terns at 23:22 local time, several hours after sunset, is possible but not typical.

The imagery also cannot exclude a fast object far from the aircraft. The parallax argument shows only that the reported speed does not require a fast-moving object, but the flicker (which does not depend on range) is the evidence that favors the bird hypothesis. A tumbling or rotating object could also flicker periodically; however, six such objects, moving together while each beats at its own steady rate of 7--8~Hz, are a less simple explanation than six birds. The altitude of the aircraft and the polarity of the display, both discussed below, relate to the bird interpretation directly (Fig.~\ref{fig:height}).

Weather-radar profiles show nocturnal migration mostly within the lowest few kilometers, reaching only 1.5~km on some nights and above 3~km on others, at heights chosen largely for favorable winds~\cite{dokter2011}. In southern Israel, autumn migrants flew mainly below 1,500~m above sea level, where the trade winds favor southward flight, and those that continued into the day at high altitude were mainly herons and gulls~\cite{bruderer1994}. Herons and gulls cruise at 4.5~km or more~\cite{schmaljohann2009}, and some songbirds climb above 6~km on migratory flights prolonged into the day~\cite{sjoberg2021}. The viewing geometry places the aircraft at 5.8~km (5.4--6.3~km) and a stationary object at 4.6~km (4.3--4.8~km). This is above the heights of typical nocturnal migration and near the cruising height of herons and gulls. It is within the heights that migrating waders are known to reach. Black-tailed godwits tracked between Europe and West Africa spent about a fifth of their migratory flights above 5~km and reached nearly 6~km, flying higher when the air at lower levels was hot~\cite{senner2018}. Great snipes cruise at about 2~km by night and 4~km by day, and one was recorded at 8.7~km~\cite{lindstrom2021}. Infrared cameras detect birds at the distances involved. A ground-based camera with a field of view of 1.45$^\circ$ detects small songbirds up to 3~km overhead in clear skies~\cite{weisshaupt2016}. A slow object here was about 3.1~km from the aircraft while the six objects were tracked and about 6.0~km during the crossing, in a field of view of 0.85$^\circ$. The crew's altitude estimate was made without a range and is subject to the same
ambiguity as the speed: an object seen along a line of sight can be at any height between the aircraft and the ground.

The height of the objects remains the greatest challenge to the bird hypothesis. The camera shows the group of objects crossing the field of view with the island and sea around it, with no horizon in view. This implies that the objects were seen against the sea surface and were therefore below the aircraft. The crew reported the objects at flight level 220 (6.7~km), which is above our estimate of the aircraft's own height, so the objects were lower than the crew judged. Even so, a slow object at 4.6~km is higher than most birds fly, and the bird hypothesis requires a small flock at a height that few species reach.

One further property of the video tests this hypothesis. At 4.6~km the standard atmosphere is at $-15\,^\circ$C, and an atmosphere 20$\,^\circ$C warmer is still near freezing, so the plumage of a bird flying there is tens of degrees colder than the Gulf, which was at 33$\,^\circ$C that day~\cite{huang2021}. Such birds appear bright against the sea only on a display set to black-hot, in which colder is brighter; a bird near the sea surface, or an engine or a flame, would not. The setting is under a redaction box, but the island scene indicates it (see Display polarity): the island's painted lattice tower, in air that was warmer than the sea, appears much darker than the sea behind it, so the display was black-hot. The six objects, bright in the same frames, were therefore colder than the sea, as birds at the height found here would be. The test could have failed: on a white-hot display they would have been warmer than the sea. With either setting, no object shows a plume, a trail, or a second source beside it: each is a single compact point whose brightness cycles steadily.

In general, a speed obtained from airborne imagery without a range describes the ground and not the object. For DOW-UAP-PR135, the aircraft's recorded altitude, track, and ground speed, the display's recorded setting, or the telemetry recorded with the video would replace the estimates made here with measurements. But the brightness flicker test operates even when range is unknown and can be performed on the released pixels of any video in which an object is tracked for a few seconds. Other analyses of UAP could also search for steady beat frequencies as evidence of flapping flight. More complete sensor and platform data in future PURSUE releases would allow competing explanations to be tested more decisively.

This single-camera video analysis has limits. The range to the objects was not measured, so their height, speed, and size are estimates. These bounds rest on one island scene. The field of view from the crossing also assumes that the objects did not accelerate during its three seconds; the island's growth checks it independently, but only to within about a fifth. The six objects were followed without a break for only 5~s. Their cores are clipped and the display is contrast-enhanced, so the shape of each brightness cycle is not preserved. At 29.97 frames per second a rhythm faster than 15~Hz would fold into the measured band (see Beat frequencies), so the frequency alone does not exclude a faster rhythm, such as a propeller. The display's polarity is inferred from the scene and from weather records, not read from the display.

This case shows the value of independent scientific investigation of UAP. Reports from the field are perceptual estimates, made in real time, during a mission, and often without the information needed to interpret what was seen. Once recorded, such estimates can define a case as anomalous. This study involved researchers in academia working only from the released sensor videos, which led to a novel measurement technique that found a simpler explanation. Military assessment serves operational needs; scientific analysis, conducted in parallel, can test competing explanations transparently and resolve cases that would otherwise remain open. What limits such analysis is not the phenomena but the data.

\section{Methods}\label{sec:methods}

\subsection{Data}\label{sec:data}

The video analyzed here is DOW-UAP-PR135, released by the U.S.\ Department of War in the sixth tranche of its PURSUE records on 18 September 2026 and hosted on the Defense Visual Information Distribution Service (DVIDS) as item 1023397~\cite{pursue,pr135video}. The file is an H.264 (High profile) stream of 1280$\times$720 pixels at 30000/1001 frames per second (29.97~Hz), 1798 frames (60.0~s), at an average bit rate of 2.93~Mb/s. We use the exact frame rate throughout this analysis, so frame $n$ is at time $t = (n-1)/29.97$~s. According to the accompanying video description~\cite{pr135video}, ``several areas of contrast transit the sensor field-of-view'' from 0:05 to 0:10, and the ``sensor pans to track the areas of contrast, keeping them generally within the center of the frame'' from 0:11 to 0:47. All of the measurements below come from the first 47 seconds. An unmodified copy of the file is archived with this paper (see Data availability), so the analysis does not depend on the public copy staying online or unchanged.

The accompanying mission report, DOW-UAP-D108~\cite{d108}, was filed by the 65th Special Operations Squadron for a contact at 2022Z on 29 July 2025 in the Middle East. Every kinematic field describing the objects in the report is marked as estimated, including a ``kinetic velocity'' of 480~mph, a southeast trajectory, and an altitude of FL220. The report's text reads ``48OMPH'', with a letter O; we interpret this as 480~mph, which is consistent with the official video description~\cite{pr135video}; however, a speed of 48~mph (21~m/s) would itself be more typical of bird flight. The report does not record that the range to the objects was measured. It lists the aircraft's own trajectory as northwest at 120 knots indicated airspeed, but the aircraft's altitude is redacted.

The report's location fields are redacted, but the first 10~s of the video show a small, low island carrying a lattice tower and a building. This is Qaruh Island, a small sandy island off the coast of Kuwait in the northern Persian Gulf (28.82$^\circ$N, 48.78$^\circ$E): its tower, buildings, and pier fall on their mapped positions (see Viewing geometry). The site tells us which kinds of birds are typically observed in the region. In local time (UTC+3), the contact time of 2022Z is 23:22, 4.7~h after sunset. The Moon was 25\% illuminated and had set at 21:26 local time~\cite{astropy2022}, so the sky was dark.

The aircraft type is also redacted. We infer that it was an MQ-9 Reaper, a remotely piloted aircraft, from three entries in the report. First, the 65th Special Operations Squadron operates the MQ-9~\cite{sos65}. Second, a single aircraft flew for 20.6~h (takeoff 1734Z on 29 July, landing 1411Z on 30 July). Third, its speed of 120 knots is a normal MQ-9 cruise speed. This would indicate that the sensor video was taken while the crew operated the aircraft from a remote location. If so, the thermal camera of its sensor turret most likely records mid-wave infrared light (3--5~$\mu$m), not the long-wave infrared light of the Galileo Project's cameras (see Infrared video of flying animals).
However, none of our results depends on the aircraft type. The viewing geometry takes the field of view from the video itself and not from the specifications of any sensor (see Viewing geometry). The ground-projected speed analysis (equation~\eqref{eq:vg}) uses only the aircraft's reported speed and track, and assumes that no range measurement of the objects was taken.

\subsection{Frame decoding and conventions}

Every frame is decoded losslessly to an 8-bit image with FFmpeg and converted to a single luminance channel. We refer to the resulting pixel values as digital numbers (DN, 0--255). The objects appear as bright points. The part of the display that states whether it was set to white-hot or to black-hot is under a redaction box. The island scene indicates black-hot, in which colder is brighter (see Display polarity). Positions are in pixels, with $x$ to the right and $y$ down. All rates are measured relative to the video frame, in screen coordinates; we explicitly state whenever we measure a rate relative to something else, such as the background or the ground. Redaction boxes (large regions of zero value) and the burned-in symbology are masked before any analysis.

\subsection{Group motion}

The group's rate across the held scene is measured on frames 187--280, against the island (see Viewing geometry). The azimuth of the line of sight is read in every frame from the burned-in north pointer, which the display draws at a fixed radius from the boresight reticle. We locate the pointer glyph by template matching and take its angle about the boresight. A turn of this pointer measures a change in the sensor's azimuth.

\subsection{\texorpdfstring{Viewing geometry}{Viewing geometry}}\label{sec:geometry}

\emph{The island.} The island held in view from 0:00 to 0:10 was identified from its layout. The lattice tower, the buildings beside it and to its northwest, and the pier that runs southwest from the shore all fall on their positions on the map of Qaruh Island~\cite{osm} when the island is viewed from the southeast. The mapped outlines of the tower's base (12.5 by 13.0~m), of two buildings with bright roofs in the video, of two shelters beside the tower, and of the pier serve as control features. The shore is not used, because the shore of a sand island is not a constant boundary. Drawn with the fitted camera, the mapped shore agrees with the picture on the far and near sides of the island and lies up to about 20~m from it at the two ends (Fig.~\ref{fig:scene}, top).

\emph{The averaged picture.} The island's shift on the screen is measured on every frame from 1 to 300 by normalized cross-correlation of the gradient magnitude of a patch of 100 by 190 pixels that contains the tower. While it is held, the scene drifts by 9.0 pixels per second to the left and 3.0 pixels per second upward. Frames 1--180 are shifted back by these amounts and averaged. This sharpens the island and smears out the sensor's fixed pattern, which does not move with the scene. On the averaged picture we read the tower's axis and its lean, the outer edges of its legs just above the base, the outline of the bright roofs at half their brightness above the surroundings, the line of the pier, and the left edge of the shelters.

\emph{The camera.} The field of view is under one degree, so perspective across the island is negligible and the camera is modeled as a weak-perspective projection. A point $E$ meters east and $N$ meters north of the tower's base, and $U$ meters above the ground there, appears at
\begin{equation}
x = x_0 + \frac{E\cos\alpha - N\sin\alpha}{s}, \qquad
y = y_0 - \frac{(E\sin\alpha + N\cos\alpha)\sin\varepsilon + U\cos\varepsilon}{s},
\label{eq:camera}
\end{equation}
followed by a rotation of the picture about $(x_0, y_0)$ by a small roll angle. Here $s$ is the scale in meters per pixel across the line of sight at the island, $\varepsilon$ the angle of the line of sight below the horizontal, and $\alpha$ its azimuth. These three, the roll, the position $(x_0, y_0)$ of the tower's base, the height of the roofs, and the height of the pier's deck are found by least squares. The look-down angle is fixed mainly by the depth of the two roofs on the picture, which lie at one height, and the scale by their width and by the tower's base. The fit has two nearly equal minima, 1.5$^\circ$ apart in look-down angle, so it is started from several points. Uncertainties come from 1,000 repetitions of the fit in which the inputs are redrawn: the map's scale (2\%), the position of each structure (1~m), the size of the roofs (4\%) and of the tower's base (8\%), each reading (1--3 pixels), the brightness level at which the roofs' edge is read, and the heights assumed beforehand for the roofs ($4.0 \pm 1.5$~m) and the pier's deck ($0 \pm 1$~m). We quote the median and the range that holds the middle 68\% of the 1,000 fits (16th to 84th percentiles). The fit gives $s = 0.32$~m per pixel (0.31--0.33), $\varepsilon = 12.1^\circ$ (11.4--12.9$^\circ$), and $\alpha = 325^\circ$ (323--326$^\circ$). The azimuth is 5--7$^\circ$ greater than the direction shown by the display's north pointer (318--320$^\circ$). The local magnetic declination, 3.6$^\circ$ east, accounts for most of this difference if the pointer marks magnetic north. The camera assumes square pixels, which the tower confirms. On the fitted camera the tower is 86~m tall (see Display polarity). On a satellite image of the island taken on 21 January 2022 (WorldView-2, Vantor, in Esri World Imagery), its shadow runs 115~m from the center of its base toward azimuth 334$^\circ$. On that day the sun stood at the opposite azimuth at 10:27 local time, 37$^\circ$ above the horizon, so the tower is 86~m tall (80--91~m). Had the sensor's 4:3 picture been stretched to fill the 16:9 frame, the fitted camera would have made it 117~m tall.

\emph{The crossing.} Each frame from 176 to 296 is compared with the averaged picture shifted to that frame, and the sensor's fixed pattern, taken as the median of these differences, is removed. Bright points are the local maxima of what remains. On each of 82 frames from 187 to 280, the mean position of the leading row of three to six points is expressed in the coordinates of the scene. A parabola fitted to these positions down the screen gives the rate, 198.1 pixels per second, and its growth, 6.5 pixels per second squared (6.3--6.8, from residuals resampled in half-second blocks). The scatter is 0.55 pixel about the parabola and 2.4 pixels about a straight line. Across the screen the rate is 69.0 pixels per second to the left, with no significant change. A rate $v_y$ down the screen is a speed $v_y s/\sin\varepsilon$ along the line of sight on the sea surface, and a rate $v_x$ across the screen is a speed $v_x s$ across it. The width of each point is its full width at half maximum across the row: 3.9 pixels, about the width of the sensor's own blur. The leading row spans 29 pixels, with its members about 7 pixels apart.

\emph{Field of view and heights.} The group's rate down the screen grows because the aircraft approaches. Let $\omega$ be the rate, in radians per second, at which the line of sight to an object turns downward. Suppose the aircraft closes on the object at a steady relative speed $V$ that is level and directed along the line of sight in plan view, as the steady north pointer indicates (see Group motion). If the object is at slant range $\rho$, the line of sight turns downward at $\omega = V\sin\varepsilon/\rho$, and differentiating gives
\begin{equation}
\frac{d\omega}{dt} = \frac{2\,\omega^{2}}{\tan\varepsilon},
\label{eq:rate}
\end{equation}
in which neither $V$ nor $\rho$ appears. The group is measured against the island, which the sensor holds, so its rate and the growth of that rate are the object's less the island's. In pixels, the island's line of sight turns at $w_{\rm i} = v_{\rm air}\sin\varepsilon/s$ pixels per second, and by equation~\eqref{eq:rate} that rate grows by $2w_{\rm i}^{2}\beta/\tan\varepsilon$ pixels per second squared, where $\beta$ is the angle that one pixel spans. For $v_{\rm air}$ we take the true airspeed (equation~\eqref{eq:tas}) with an unknown wind of $\pm$10~m/s along the track. The true airspeed depends on the aircraft's height, which follows from $\beta$, so the calculation is repeated until it converges. With the group's rate $\dot{y}$ and its growth $\ddot{y}$ down the screen, the object's rate is $w = \dot{y} + w_{\rm i}$, and equation~\eqref{eq:rate} gives $\ddot{y} + 2w_{\rm i}^{2}\beta/\tan\varepsilon = 2w^{2}\beta/\tan\varepsilon$, so that $\beta = \ddot{y}\tan\varepsilon/[2(w^{2}-w_{\rm i}^{2})]$, and the field of view across the frame is $1280\,\beta$. The time until the aircraft would pass over the object is $2\cos^{2}\!\varepsilon\;\omega/(d\omega/dt)$, 71~s after the crossing (68--74~s). The range to the island is $s/\beta$, and the aircraft's height is $(s/\beta)\sin\varepsilon$. The camera is fitted to the averaged picture, whose mean time is 4.7~s before that of the crossing. By the crossing the island is 1.4\% nearer and its line of sight 0.17$^\circ$ steeper, and $s$ and $\varepsilon$ are carried forward by these amounts before they are applied to the crossing. None of these steps uses the specifications of the sensor. The main assumption is that the objects themselves did not accelerate during the three seconds of the crossing: an acceleration of 0.05~m/s$^2$ down the screen would change the field of view, and with it the range and the heights, by about 10\%. The curvature is not steady over the crossing: its first and second halves, fitted separately, give 2.3 and 8.3 pixels per second squared, so the objects' own motion contributes to it.

\emph{The island's growth.} The island gives a check that assumes nothing about the objects. As the aircraft closes, the island grows on the screen at a rate $v_{\rm air}\cos\varepsilon/R$, where $R$ is its range. Frames 1--300 are averaged in ten groups of 30 frames, in the same way as the averaged picture. The tower above the far shore, through high-pass filters of three widths, is matched between groups by a change of scale, a shift, and a gain, against the first group and against the last. Its height in pixels grows by 0.30\% per second (0.28--0.32\%). With the true airspeed and the same allowance for the wind, this places the island 27~km from the aircraft (22--33~km), for a field of view of 0.87$^\circ$ (0.72--1.09$^\circ$), in agreement with the values from the crossing.

\subsection{\texorpdfstring{Display polarity}{Display polarity}}\label{sec:polarity}

Sand, beach, and sea are within a few degrees of each other at night, and the sign of their differences depends on moisture and emissivity, so they do not decide the setting of the display. The island's lattice tower does. It is painted in horizontal red and white bands~\cite{rowlett2025}, so its surface emits nearly as a black body and shows its own temperature, which for thin members in a wind is within a few degrees of that of the air. The mission report gives the time of the contact as 2022Z (20:22~UTC) on 29 July 2025 (see Data), so the weather records used here are the hourly observations at 20:00 and 21:00~UTC, on either side of that time, and the sea surface temperature of that day. At those hours Kuwait International Airport (station 40582099999 of the Integrated Surface Database~\cite{smith2011}), about 90~km from the island in the direction from which the wind blew, recorded an air temperature of 42.0--43.0$\,^\circ$C, a dew point of 7$\,^\circ$C, and a wind from 300--320$^\circ$ at 5--8~m/s. The sea surface near the island was at 33.3$\,^\circ$C on that day, in NOAA's daily Optimum Interpolation analysis, version 2.1~\cite{huang2021}. Air carried from the land over a cooler sea is cooled from below, so the air around the tower was not colder than the sea. At the island itself, the ERA5 reanalysis~\cite{hersbach2020} gives the air 2~m above the sea at 38.1 and 37.7$\,^\circ$C at 20:00 and 21:00~UTC, about 5$\,^\circ$C warmer than the sea. On the averaged picture of frames 1--180 (see Viewing geometry), the tower's outline is found row by row as the pixels that are darker than the sea on both sides of it by more than 25~DN. It rises 86~m above its base, as the fitted camera measures it, and above 28~m the mapped far shore, drawn with the same camera, places the sea behind it. In bands of up to 20 rows from 29~m to the top, the tower's darkest members are at 4--43~DN, while the darkest 1\% of the sea on either side lies at 124--167~DN. A body warmer than its background that appears darker than it places the display in black-hot. Two compact dark points at the equipment shelters beside the tower, at 12~DN against 146~DN for the island below them, fit the same reading. On a white-hot display, the painted tower would have to be colder than the sea while standing in air that was not. The inference rests on the paint, which is known from one published description, and on the air's temperature at the island, which comes from a reanalysis and from a station on the coast, not from a measurement on the island.

\subsection{Ground-projected speed}\label{sec:parallax}

Consider an aircraft at height $h_{\rm air}$ above the ground and an object at height $h_{\rm obj}$ on the same line of sight, with velocities $v_{\rm air}$ and $v_{\rm obj}$ taken as components along the aircraft's ground track. The observed motion of the ground from the aircraft, $v_{\rm gnd}$, compares to the motion of the object from the aircraft as $(v_{\rm gnd} - v_{\rm air})/h_{\rm air} = (v_{\rm obj} - v_{\rm air})/(h_{\rm air}-h_{\rm obj})$. This gives the height of the object as
\begin{equation}
h_{\rm obj} = h_{\rm air} \left( 1 - \frac{v_{\rm obj}-v_{\rm air}}{v_{\rm gnd}-v_{\rm air}} \right).
\label{eq:vg}
\end{equation}
Even a stationary object ($v_{\rm obj} = 0$) produces a non-zero ground-projected speed,
\begin{equation}
v_{\rm gnd} = -\frac{h_{\rm obj}}{h_{\rm air}-h_{\rm obj}}\,v_{\rm air},
\label{eq:stationary}
\end{equation}
which is directed opposite to the aircraft's track.
Equation~\eqref{eq:vg} assumes flat ground at sea level, which holds over the Gulf, and heights that stay constant during the measurement. The aircraft's true airspeed $v_{\rm air}$ is computed from its indicated airspeed, $v_{\rm IAS} = 120$~knots (61.7~m/s), as a function of the unknown aircraft height $h_{\rm air}$. An airspeed indicator responds to dynamic pressure, and the relationship can be obtained with the International Standard Atmosphere below 11~km,
\begin{equation}
v_{\rm air} = \frac{v_{\rm IAS}}{\sqrt{\sigma}}, \qquad
\sigma = \left(1 - \frac{L\,h_{\rm air}}{T_0}\right)^{4.256},
\label{eq:tas}
\end{equation}
where $\sigma$ is the air density relative to sea level, $T_0 = 288.15$~K is the sea-level temperature and $L = 0.0065$~K/m is the temperature lapse rate. For example, $v_{\rm air} = 75.5$~m/s at $h_{\rm air} = 4$~km and 83~m/s at 5.8~km. Instrument errors, compressibility (small at this speed), and wind are neglected. Equation~\eqref{eq:vg} is then evaluated with $v_{\rm gnd}$ equal to the measured 665~mph and to the reported 480~mph, directed opposite to the aircraft's track (Fig.~\ref{fig:height}). The size of one pixel at an object on the line of sight is $s\,(1 - h_{\rm obj}/h_{\rm air})$. During the tracked interval, 36~s after the crossing, a stationary object is about 3.1~km from the aircraft and one pixel is about 0.04~m, so the fitted widths of 4.3--5.5 pixels (see Apparent size) are 0.16--0.20~m.

\subsection{Tracking the six objects}\label{sec:tracking}

The six objects are resolved from one another and can be followed without a break from frame 1240 to frame 1389 (41.3--46.3~s). Before this interval, pairs of objects often merge, which makes them difficult to follow. In each frame, candidate sources are the local maxima (within $7\times7$ pixels) of a difference-of-Gaussians image (standard deviations of 1.5 and 5 pixels) that exceed 12~DN. Each candidate's position is refined to sub-pixel precision by the centroid of the positive response in a $5\times5$ box. The dark ringing that the display draws around each bright object is small, so it is not mistaken for a redaction box.

Tracking starts from frame 1330, where the six objects are well separated. They are lettered A--F from left to right in that frame. The tracker proceeds forward and backward from this frame. For each frame, the displacement common to the whole group (camera motion) is estimated, based on the offset that brings the most tracked objects within 3 pixels of a candidate, and then averaged over those matches. Each object's position is predicted from this group shift plus its own recent velocity. Candidates are assigned to objects by minimum total distance~\cite{kuhn1955}, within a gate of 5 pixels, widened by 2 pixels for each consecutive frame in which an object is missed.

All six objects are detected in every frame of the interval. Their identities were checked by overlaying the tracks on the frames. Figure~\ref{fig:frames} shows the group in four consecutive frames (44.01--44.11~s, dashed lines in Fig.~\ref{fig:flicker}), spanning about one beat on one display scale, so that the rise and fall of each object's brightness can be seen directly.

\subsection{Photometry}\label{sec:photometry}

The brightness of each object is the sum over a circular measurement aperture (i.e., a region of the image, not the camera's optical aperture). This measurement aperture has a radius of 4 pixels and is centered on the object's sub-pixel position, after subtracting the median of an annulus from 7 to 10 pixels. Pixels on the aperture's edge are weighted by the fraction of their area inside the circle, computed on an $8\times8$ sub-pixel grid. The positions are smoothed with a 5-frame running mean first, so that the aperture does not follow centroid noise. A small aperture that is not weighted this way makes a moving point flicker by itself, as its image moves across the pixel grid. We found this effect with a $3\times3$ aperture, so we use a larger aperture with weighted edges.

Slow changes in brightness (from range, background, or gain) are removed by expressing the brightness $F(t)$ in the measurement aperture, at time $t$ in the clip, as a fractional deviation from its 15-frame (0.50~s) running mean $\bar{F}(t)$: $\delta(t) = F(t)/\bar{F}(t) - 1$. This brightness deviation $\delta$ is plotted in Figs.~\ref{fig:flicker} and~\ref{fig:groups}. The first and last 7 frames are dropped from the running mean, which leaves frames 1247--1382 (136 frames, 4.54~s) for each object. The running mean follows slow changes in brightness (38--61\% over the record) but changes the measured amplitude of a 7.5~Hz beat by only about 7\%. Dividing instead by the mean of the whole record, with or without a straight-line trend removed, changes the measured beat frequencies by at most 0.03~Hz. A frame is flagged when another object lies within 9 pixels, which is the distance at which the two apertures overlap. This happens to B and C in 16\% of frames.

\subsection{Beat frequencies}\label{sec:spectra}

We subtract the mean of each series $\delta(t)$ and multiply it by a raised-cosine taper that falls smoothly to zero at both ends~\cite{harris1978}, to limit leakage of power between frequencies. We then take its discrete Fourier transform $\hat{\delta}(f)$, zero-padded to 32 times its length so that the peak can be located finely. The amplitude spectrum is $2|\hat{\delta}(f)|/\sum_n w_n$, where $w_n = \sin^2[\pi n/(N-1)]$ is the weight of the taper on the $n$-th of the $N$ samples, so that a sinusoid of amplitude $a$ gives a peak of height~$a$. The beat frequency of an object is the highest peak between 1.5 and 14~Hz. The frequency resolution of a 4.54~s (136-frame) record is 29.97\,frames per second / 136 frames = 0.22~Hz, and the sampling limit is 29.97\,frames per second / 2 = 14.99~Hz. A beat at a true frequency $f$ above the sampling limit would therefore appear at $29.97 - f$. Beats of different objects are compared through their cross-spectrum, with the phase at the peak giving the objects' relative timing.

\subsection{Apparent size}\label{sec:size}

The beat frequency is also calculated from two measures of each object's apparent size (Fig.~\ref{fig:flicker}). The first is the pixel area: the number of pixels within 7 pixels of the object that are more than 60~DN above the local background, taken as the median of the annulus from 9 to 11 pixels. The second is a fitted width: an elliptical Gaussian with a constant background, fitted to a $13\times13$ pixel stamp by least squares. The width is reported as the full width at half maximum, $2\sqrt{2\ln 2}\,\sqrt{\sigma_x\sigma_y} \approx 2.355\sqrt{\sigma_x\sigma_y}$, where $\sigma_x$ and $\sigma_y$ are the fitted standard deviations. Both series are detrended in the same way and analyzed exactly as the brightness.

The objects' cores are clipped at the top of the display range (that is, the brightness in the center exceeds the maximum recorded value). The brightest pixel reaches 255~DN in 82--100\% of frames, with 4--6 pixels at that value. Clipped pixels carry no information about the true profile, so they are excluded from the fit. Clipping couples size to brightness. When the core is clipped, a brighter object can only show up as a larger clipped area. The pixel area indeed rises and falls in step with the brightness (Fig.~\ref{fig:flicker}). We therefore treat the size measures as independent estimates of the beat frequency, rather than a probe into the actual physical size of the objects.

\subsection{\texorpdfstring{Position}{Position}}\label{sec:position}

The position of each object is taken relative to the mean position of the other five, which removes the motion of the camera. It is then treated exactly as the brightness: the 0.5~s running mean is subtracted, the ends are dropped, and the amplitude spectrum of each coordinate is computed. For each object we measure the amplitude at its brightness beat frequency and the largest amplitude within one resolution element of half that frequency. At the beat the positions cycle by 0.44--0.82 pixel (the two coordinates combined), 6 to 12 times the median level of their spectra. At half the beat they cycle by 0.04--0.12 pixel, which is at the level of the noise. Dots planted with a brightness that beats by $\pm$20\% on a smooth path (see Controls) move the tracker's center by at most 0.10 pixel at the beat, so the cycle is not made by the brightness alone. The cycle is about a quarter of a period out of step with the brightness (52--103$^\circ$) and lies along the same diagonal of the frame for all six objects. We have not established why, and we rely only on where the cycle is found: at the beat frequency and not at half of it. The tracks of the five Galileo flyers were read in the same way. The position of each cycles most strongly at its wingbeat, by 0.10--0.50 pixel. This includes the flyer whose brightness is strongest at twice its wingbeat: its position cycles by 0.40 pixel at 3.9~Hz and by 0.19 pixel at 7.8~Hz. An independent reading of the six objects by the mcDonald UAP Toolkit (see Other PURSUE videos), with its own center for each object and a test against each track's own wander, finds a cycle at the brightness beat frequency in all six (0.45--0.55 pixel).

\subsection{Controls}\label{sec:controls}

We tested several instrumental effects that could produce a periodic flicker of a few hertz.

\emph{Video compression.} The video stream is encoded with three bidirectional frames between reference frames; this means the frame types repeat every fourth frame, which gives a cadence of 29.97\,frames per second/4\,frames = 7.49~Hz, close to the observed beats. We planted six dots of constant brightness over a frame in the video, moving at the same screen rates as the objects, and added the clip's own temporal noise. We encoded these modified clips with the same settings as the original video clip (H.264 High profile, a keyframe every 60 frames, 3 bidirectional frames, 2.9~Mb/s), decoded them, and measured them using the same method as with the objects. Planting synthetic objects and measuring them through the same steps is what astronomers call an injection--recovery test.

\emph{Pixel phase.} The same planted dots were measured before encoding, to test whether the measurement aperture alone causes a moving point to flicker (see Photometry).

\emph{Common cadence.} Any flicker made by the video itself (such as dropped or repeated frames, compression, or automatic gain) would affect every object at the same frequency and in the same phase. We therefore compared the frequencies of the objects and their cross-spectral phases over the same frames.

\emph{Sensor defects.} Ten fixed bright pixels of the sensor (defects that stay in one place on the screen) were tracked and measured using the same method as with the objects.

\emph{Size or brightness.} The effect of size and brightness was tested with further sets of planted dots that are matched to the objects and encoded in the same way. The first set was constructed with brightness beats of $\pm$20\% at 7.0, 7.5, and 7.9~Hz while the width is fixed: a size measure that beats on these dots is following brightness, not size. The second set was constructed with width beats of $\pm$10\% at the same frequencies while the peak is fixed: a size measure that misses these dots could not detect a real change in size. Finally, the Gaussian fit to the real objects was repeated with a fixed central disk (radius 2.5 or 3 pixels) excluded in every frame, in addition to the clipped pixels, so that the fitted region no longer grows with brightness. If the clipped pixels are excluded from the fit, then the fitted width of the objects beats in step with their brightness (within 14$^\circ$). When the same central region is excluded from the fit in every frame, so that the fitted pixels no longer depend on brightness, the width no longer beats in step with the brightness.

\emph{Position.} The dots whose brightness beats at a fixed width were also located with the tracker's own centroid, and their positions, less their true paths, were read as the objects' positions are (see Position).

\subsection{Infrared video of flying animals}\label{sec:galileo_methods}

We applied the frame-by-frame analysis method to video from an all-sky infrared camera array of the Galileo Project (a Dalek) at the project's Hideout site near Las Vegas, Nevada~\cite{loeb2026}. The Dalek holds eight
uncooled FLIR Boson~640 cameras that observe at infrared wavelengths (7.5--13.5~$\mu$m)~\cite{domine2025,loeb2026}. Each has a $50^\circ\times40^\circ$ field of view (about 1.4~mrad per pixel) centered 30$^\circ$ above the horizon. The files are 8-bit, compressed with H.265, and recorded at 60 frames per second. The sampling limit is therefore 30~Hz,
twice that of DOW-UAP-PR135. The cameras record only in daylight, so we used the first and last hours of recording, when birds are most active.

The five flyers in Fig.~\ref{fig:groups} were recorded by cameras 1 and 5 on the evenings of 19 and 22 September 2026 (17:20--18:47 local time) and on the mornings of 2 and 5 February 2026 (07:14--07:15 local time). Moving objects were found by subtracting a running average of the scene and linked into tracks. Each track was then measured exactly as the DOW-UAP-PR135 objects were (see Photometry and Beat frequencies), with the 0.5~s running mean spanning 30 frames. The beat frequencies from brightness and pixel area agree within 0.1~Hz for each flyer, and each record lasts 1.5--14.9~s. As a control, the same measurement aperture was placed 15 pixels to either side of each track, on empty
sky, where it shows 4--28\% of the flyer's amplitude at the flyer's frequency. A point of constant brightness moving across the pixel grid at each flyer's rate, measured in the same way, flickers by at most 2.0\% at unrelated frequencies, compared to 7--28\% for the flyers (see Photometry).

\subsection{\texorpdfstring{Other PURSUE videos}{Other PURSUE videos}}\label{sec:other_videos}

We applied the measurements described above (see Photometry, Beat frequencies, and Apparent size) to the other sensor videos released under PURSUE, up to its eighth release of 20 November 2026~\cite{pursue}. In addition to DOW-UAP-PR135, twelve videos (DOW-UAP-PR041, DOW-UAP-PR044, DOW-UAP-PR077, DOW-UAP-PR079, DOW-UAP-PR085, DOW-UAP-PR088, DOW-UAP-PR089, DOW-UAP-PR097, DOW-UAP-PR099, DOW-UAP-PR141, DOW-UAP-PR168, and DOW-UAP-PR169) show objects with brightness beats, at 2.3--11.8~Hz, with the pixel area beating in step with the brightness (Supplementary Figs.~S1--S14). Table~\ref{tab:beats} lists every beating object, with its beat frequency and the frames analyzed. Each object has a letter of its own, continuing from the six objects A--F of DOW-UAP-PR135. DOW-UAP-PR097 holds, besides the four points flying together (N--Q), four more beating objects at other times in the clip (Supplementary Fig.~S9): a bright knot of two or three points at 6.80~Hz (R), two dark points in the frame together at 7.45 and 6.63~Hz (S and T), and a dark point at 2.29~Hz (U). In DOW-UAP-PR088 the object that beats at about 10.5~Hz (M) was followed by hand for 118~s (frames 5300--8857): its beat is present in 53 of 58 four-second windows, its rate wanders between 10.0 and 11.2~Hz, and no pause in the beat lasts longer than one second. DOW-UAP-PR089 is the second part of the same recording. Its first frame shows the scene of the last frame of DOW-UAP-PR088 to within about 3~pixels, and the object lies 89~pixels further along its path, as it would after about one second. The object beats there at 10.51~Hz for 45~s more (frames 1--1350), less steadily than before: its beat is present in 11 of 21 four-second windows. We count it as one object, M, in the two videos. In DOW-UAP-PR168, four points in view together at 4--12~s beat at 7.04--8.73~Hz (Z--AC, Supplementary Fig.~S12), and three more, in view together at 14--16~s, beat at 7.86--8.48~Hz over about one second each (AD--AF, Supplementary Fig.~S13). These three appear after the first four have left the view, so some of them could be the same objects seen again. The object in DOW-UAP-PR169 (AG, Supplementary Fig.~S14) beats most strongly at 11.77~Hz and also at half that rate, 5.9~Hz, where its position cycles, by 0.6~pixel. The position of the Galileo flyer whose brightness is strongest at twice its wingbeat cycles in the same way, at the wingbeat (see Position), so this object's wingbeat would be 5.9~Hz. These rates lie within the wingbeat frequencies of birds in cruising flight~\cite{pennycuick2001,bruderer2010} and close to the 3.4--10.0~Hz of the flying animals in Fig.~\ref{fig:groups}. As with DOW-UAP-PR135, these videos give no range to the objects, so their sizes and speeds cannot be measured, and a tumbling body could also beat at these rates. For each object, the background beside it does not beat, and the beat is not a repeat cadence of the encoding. All twelve videos are H.264 streams. The ten from the first six releases run at exactly 30 frames per second, and DOW-UAP-PR169 at 29.97, as DOW-UAP-PR135 does. DOW-UAP-PR168 is stored at 60 frames per second but shows every picture twice, so it was measured on its 30 new pictures per second; its frames in Table~\ref{tab:beats} are numbered as in the file.

The measurements of all thirteen videos were repeated with the mcDonald UAP Toolkit, version 0.5.8~\cite{mcdonald}, a deterministic and open-source package that carries out this analysis on any video and writes each object's tracks, measurements, figures, and conclusion to one case file (see Code availability). The toolkit applies a stricter test than the one above. A beat must stand out at the 1\% level against noise shaped like the object's own brightness spectrum, over the whole track or in two-second windows that agree in frequency, and it must reach 7\% of the object's brightness, a floor set on labeled infrared video of birds, aircraft, and drones~\cite{svanstrom2021}. By this test all six objects of DOW-UAP-PR135 beat like a bird, at the frequencies of Table~\ref{tab:beats} to within 0.01~Hz, and so do 17 of the 27 objects in the other videos, to within 0.35~Hz. The other ten (K, O, Q, T, W, Y, Z, AC, AD, and AE, in DOW-UAP-PR085, DOW-UAP-PR097, DOW-UAP-PR099, DOW-UAP-PR141, and DOW-UAP-PR168) are too faint or do not stand clear of the noise at that level; the last column of Table~\ref{tab:beats} marks them, and they should be read as candidates. Object M passes in DOW-UAP-PR088. Over its 45~s in DOW-UAP-PR089 its beat stands clear of the noise, but it reaches only 6\% of the object's brightness, just under the floor. The toolkit reads DOW-UAP-PR168 as stored, so each picture counts twice in its tests, as it does in the noise, background, and planted dots that they compare the object with.

DOW-UAP-PR088 and DOW-UAP-PR089 are paired with a report, DOW-UAP-D042~\cite{d42}, a debrief form dated 31 August 2020. It describes three objects on the screen at once, ``moving amongst each other'', and gives the contact an altitude of 18,000~ft and a speed of 230 knots. The form records no radar track and does not say how either number was obtained. As with DOW-UAP-PR135, a camera alone gives no range from which to measure them.

\begin{table}[h]
\caption{Brightness beats of objects in thirteen of the PURSUE videos. DOW-UAP-PR089 continues the recording of DOW-UAP-PR088, and its object is object M of that video. The object of DOW-UAP-PR169 beats at two rates. The last column says whether the object passes the stricter test of the mcDonald toolkit (see Other PURSUE videos).}\label{tab:beats}%
\begin{tabular}{@{}lllll@{}}
\toprule
Video & Object & Beat frequency (Hz) & Frames analyzed & Stricter test \\
\midrule
DOW-UAP-PR135 & A & 7.85 & 1247--1382 & Yes \\
 & B & 7.38 & 1247--1382 & Yes \\
 & C & 7.85 & 1247--1382 & Yes \\
 & D & 7.11 & 1247--1382 & Yes \\
 & E & 7.64 & 1247--1382 & Yes \\
 & F & 7.60 & 1247--1382 & Yes \\
DOW-UAP-PR041 & G & 6.43 & 2100--2836 & Yes \\
DOW-UAP-PR044 & H & 3.06 & 8040--8340 & Yes \\
DOW-UAP-PR077 & I & 9.93 & 6263--6567, 6760--6930 & Yes \\
DOW-UAP-PR079 & J & 2.34 & 831--1198 & Yes \\
DOW-UAP-PR085 & K & 7.37 & 2983--3923 & No \\
DOW-UAP-PR088 & L & 3.1--3.5 & 3883--4208, 4650--4951 & Yes \\
 & M & 10.2--11.0 & 5067--8963 & Yes \\
DOW-UAP-PR089 & M (continued) & 10.51 & 1--1350 & No \\
DOW-UAP-PR097 & N & 6.86 & 3755--3880 & Yes \\
 & O & 6.95 & 3755--3880 & No \\
 & P & 6.85 & 3755--3880 & Yes \\
 & Q & 6.90 & 3755--3880 & No \\
 & R & 6.80 & 4579--4682 & Yes \\
 & S & 7.45 & 6830--6986 & Yes \\
 & T & 6.63 & 6780--7061 & No \\
 & U & 2.29 & 7232--7413 & Yes \\
DOW-UAP-PR099 & V & 5.97--6.13 & 2781--4105 & Yes \\
 & W & 6.70 & 3257--3390 & No \\
 & X & 4.84--4.99 & 4427--8731 & Yes \\
DOW-UAP-PR141 & Y & 8.24--8.32 & 2346--2440 & No \\
DOW-UAP-PR168 & Z & 8.25 & 245--713 & No \\
 & AA & 8.29 & 251--461, 475--725 & Yes \\
 & AB & 8.73 & 247--473, 485--705 & Yes \\
 & AC & 7.04 & 281--475, 495--651 & No \\
 & AD & 8.48 & 857--929 & No \\
 & AE & 8.17 & 861--943 & No \\
 & AF & 7.86 & 875--943 & Yes \\
DOW-UAP-PR169 & AG & 11.77 and 5.9 & 36--143, 197--403 & Yes \\
\botrule
\end{tabular}
\end{table}

The analysis code, the figures, and drafts of the text were developed with Claude (Anthropic; the Opus and Fable models), a large language model.

\backmatter

\bmhead{Data availability}

Unmodified copies of the video file DOW-UAP-PR135 as released, of the mission report DOW-UAP-D108, and of the twelve other PURSUE videos of Table~\ref{tab:beats} are archived on Zenodo (\url{https://doi.org/10.5281/zenodo.24406462}), with a SHA-256 checksum for every file, so that the analysis does not depend on the public copies~\cite{pursue,pr135video,d108} remaining online or unchanged. The infrared footage of flying animals was provided by the Galileo Project~\cite{domine2025,loeb2026}. Short clips of the five flyers in Fig.~\ref{fig:groups} were cut from the original files without re-encoding and are in the same archive, with the map data of Qaruh Island~\cite{osm}, the reading of the tower's shadow on a satellite image, the weather records used for the display polarity~\cite{smith2011,huang2021,hersbach2020}, and every table of results behind the figures.

\bmhead{Code availability}

All code is archived with the data on Zenodo (\url{https://doi.org/10.5281/zenodo.24406462}). Stand-alone scripts allow for replication of every measurement of DOW-UAP-PR135 and of the Galileo flyers from the video files, along with all figures. The archive also includes a case file for each of the thirteen videos of Table~\ref{tab:beats}, made with the open-source mcDonald UAP Toolkit, version 0.5.8~\cite{mcdonald} (\url{https://doi.org/10.5281/zenodo.23652954}). A case file opens in the toolkit, which repeats every measurement in it from the video with one command.

\bibliography{refs}

\bmhead{Acknowledgements}
The authors conducted this study as part of the UAP Science Advisory Council. We thank the Galileo Project for providing the infrared footage from its Dalek camera array at the Hideout site.

\bmhead{Author contributions}
J.H.M. conceived the study, performed the analysis of the PURSUE videos and the Galileo recordings, prepared the figures, and wrote the manuscript. R.K. contributed to the testing of analysis tooling and interpretation of results. R.C. and A.L. provided the infrared recordings from the Galileo Project. O.E. and M.S. contributed to the manuscript. All authors reviewed the manuscript.

\bmhead{Funding}
J.H.M. acknowledges support from Emergent Ventures at the Mercatus Center.

\bmhead{Competing interests}
The authors declare the following competing interests: All authors are unpaid members of the UAP Science Advisory Council, under which this study was conducted. The authors declare no financial competing interests.

\end{document}


\begin{center}
{\Large\bfseries Supplementary Information}\\[8pt]
{\large Six objects reported at 480 mph in a declassified military infrared video show brightness
flicker at bird wingbeat rates}\\[8pt]
Jacob Haqq-Misra$^{1,*}$, Ravi Kopparapu$^{2}$, Richard Cloete$^{3}$, Omer Eldadi$^{3}$,
Michael Shermer$^{4}$, Abraham Loeb$^{3}$\\[4pt]
{\small $^1$Blue Marble Space, Seattle, WA, USA;
$^2$NASA Goddard Space Flight Center, Greenbelt, MD, USA;
$^3$Galileo Project, Harvard University, Cambridge, MA, USA;
$^4$The Skeptics Society, Santa Barbara, CA, USA.
$^*$jacob@bmsis.org}
\end{center}

\begin{figure}[!htbp]
\centering
\includegraphics[width=\textwidth]{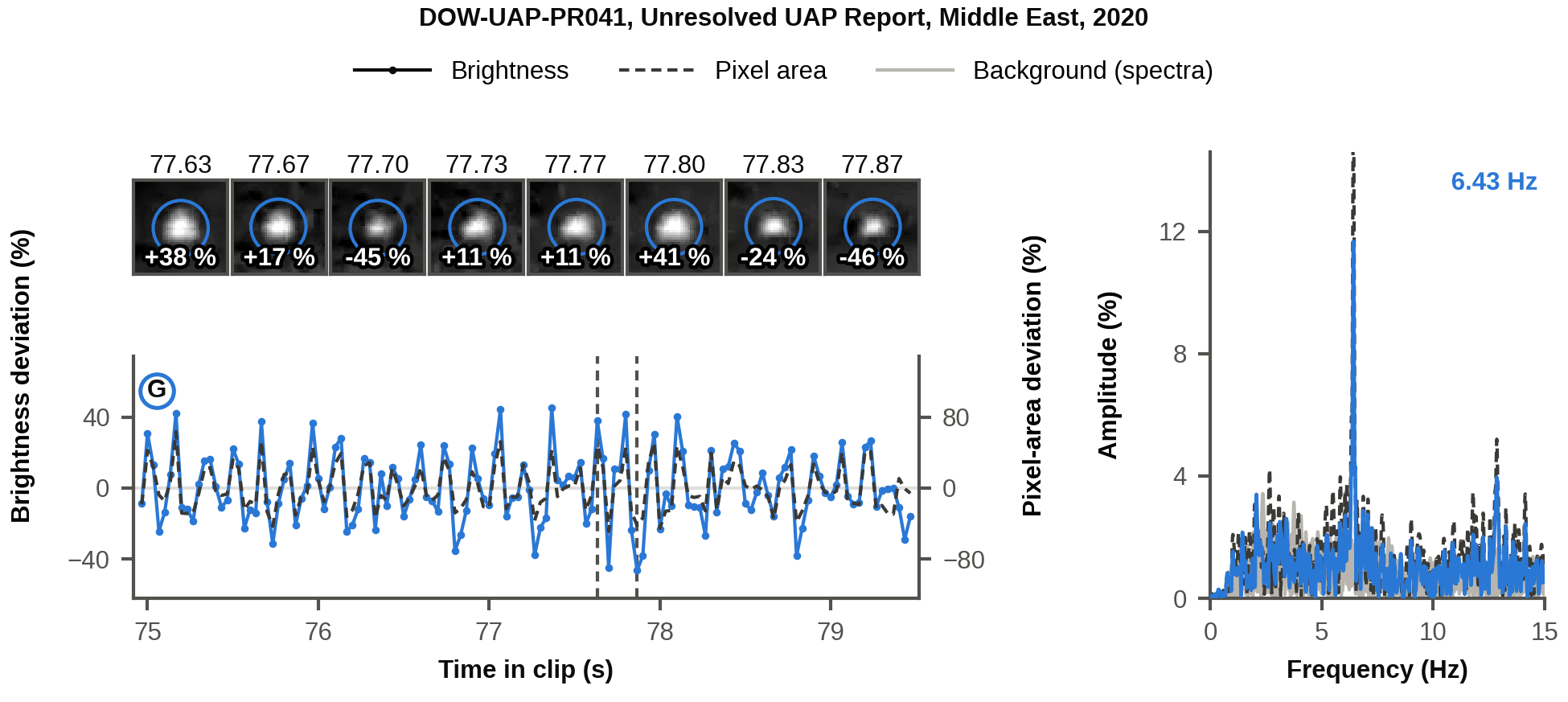}
\caption{\rev{Object G, a bright point over water in DOW-UAP-PR041, beats at 6.43~Hz. Objects are lettered as in Table~1 of the main text. Video DOW-UAP-PR041, ``Unresolved UAP Report, Middle East, 2020''~\cite{pr41video}.}}
\label{fig:pr41}
\end{figure}

\begin{figure}[!htbp]
\centering
\includegraphics[width=\textwidth]{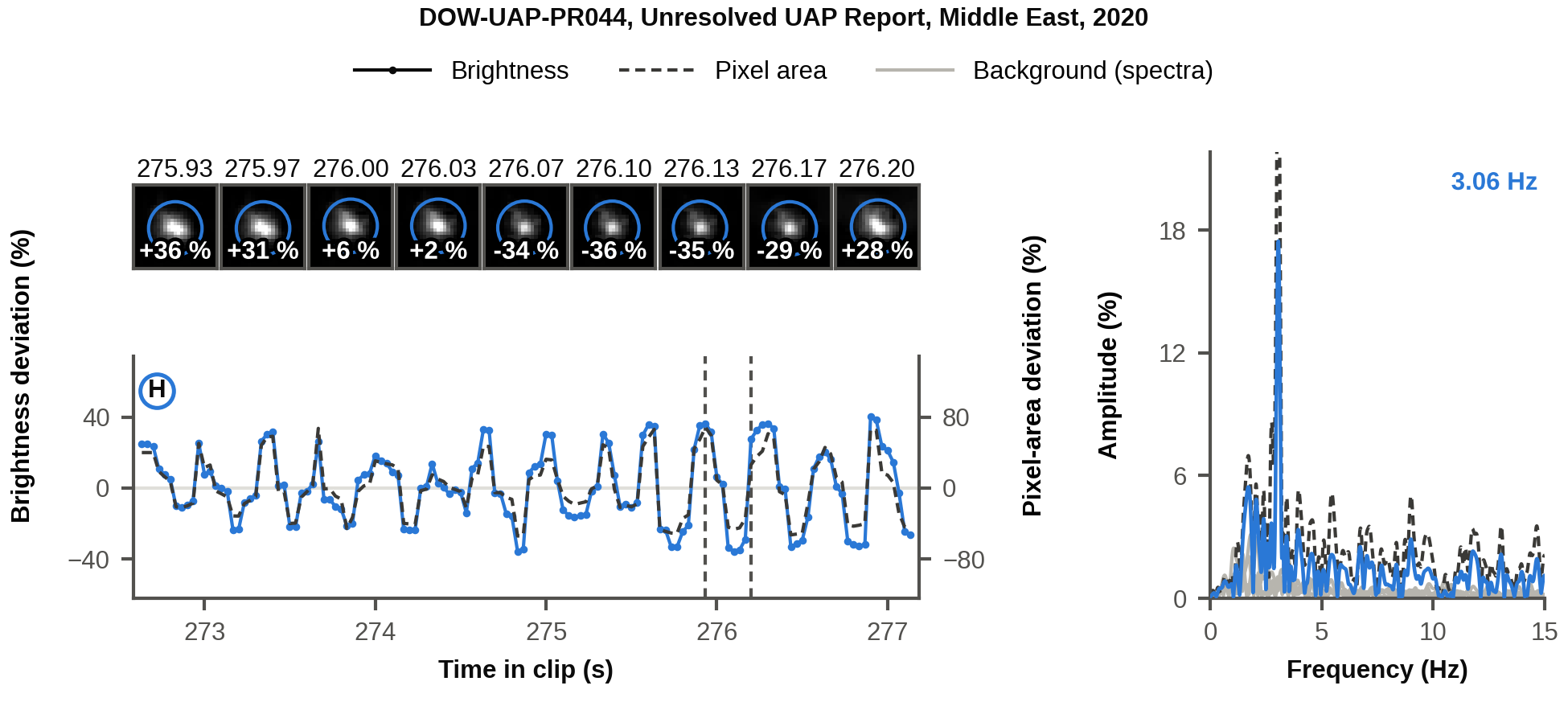}
\caption{\rev{Object H, a bright point held over water in DOW-UAP-PR044, beats at 3.06~Hz. Video DOW-UAP-PR044, ``Unresolved UAP Report, Middle East, 2020''~\cite{pr44video}.}}
\label{fig:pr44}
\end{figure}

\begin{figure}[!htbp]
\centering
\includegraphics[width=\textwidth]{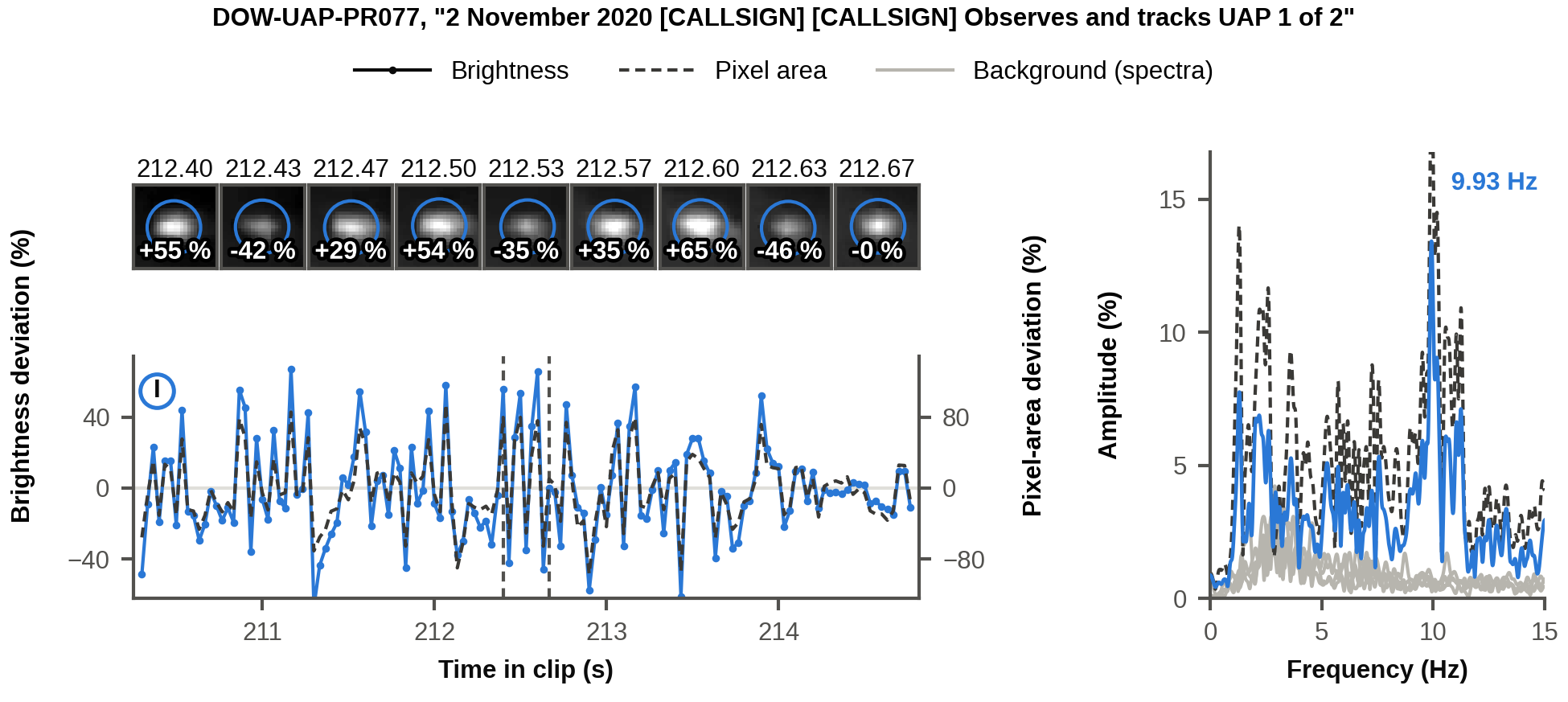}
\caption{\rev{Object I, a bright point over cloud in DOW-UAP-PR077, beats at 9.93~Hz. Video DOW-UAP-PR077, ``2 November 2020 [CALLSIGN] [CALLSIGN] Observes and tracks UAP 1 of 2''~\cite{pr077video}.}}
\label{fig:pr077}
\end{figure}

\begin{figure}[!htbp]
\centering
\includegraphics[width=\textwidth]{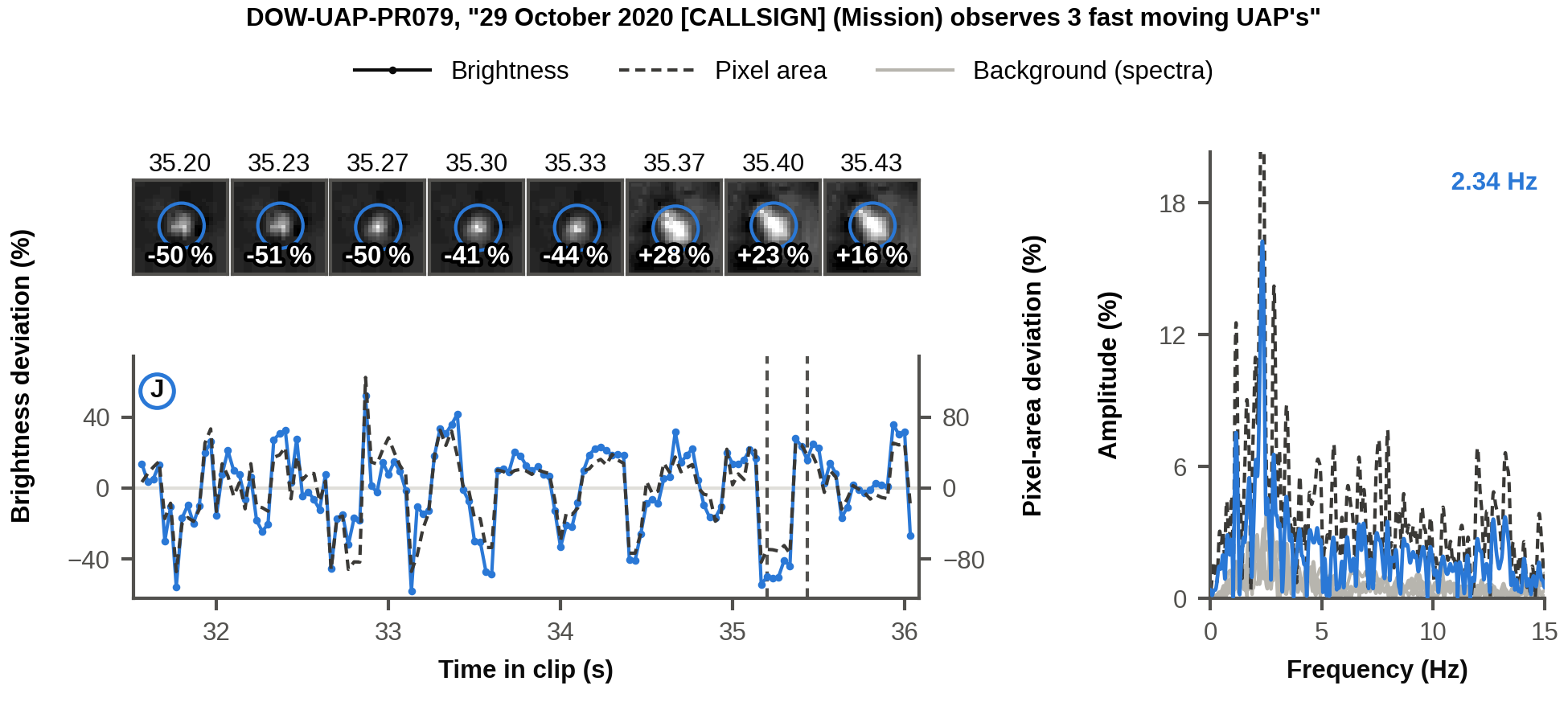}
\caption{\rev{Object J, a point in DOW-UAP-PR079, beats at 2.34~Hz. Video DOW-UAP-PR079, ``29 October 2020 [CALLSIGN] (Mission) observes 3 fast moving UAP's''~\cite{pr079video}.}}
\label{fig:pr079}
\end{figure}

\begin{figure}[!htbp]
\centering
\includegraphics[width=\textwidth]{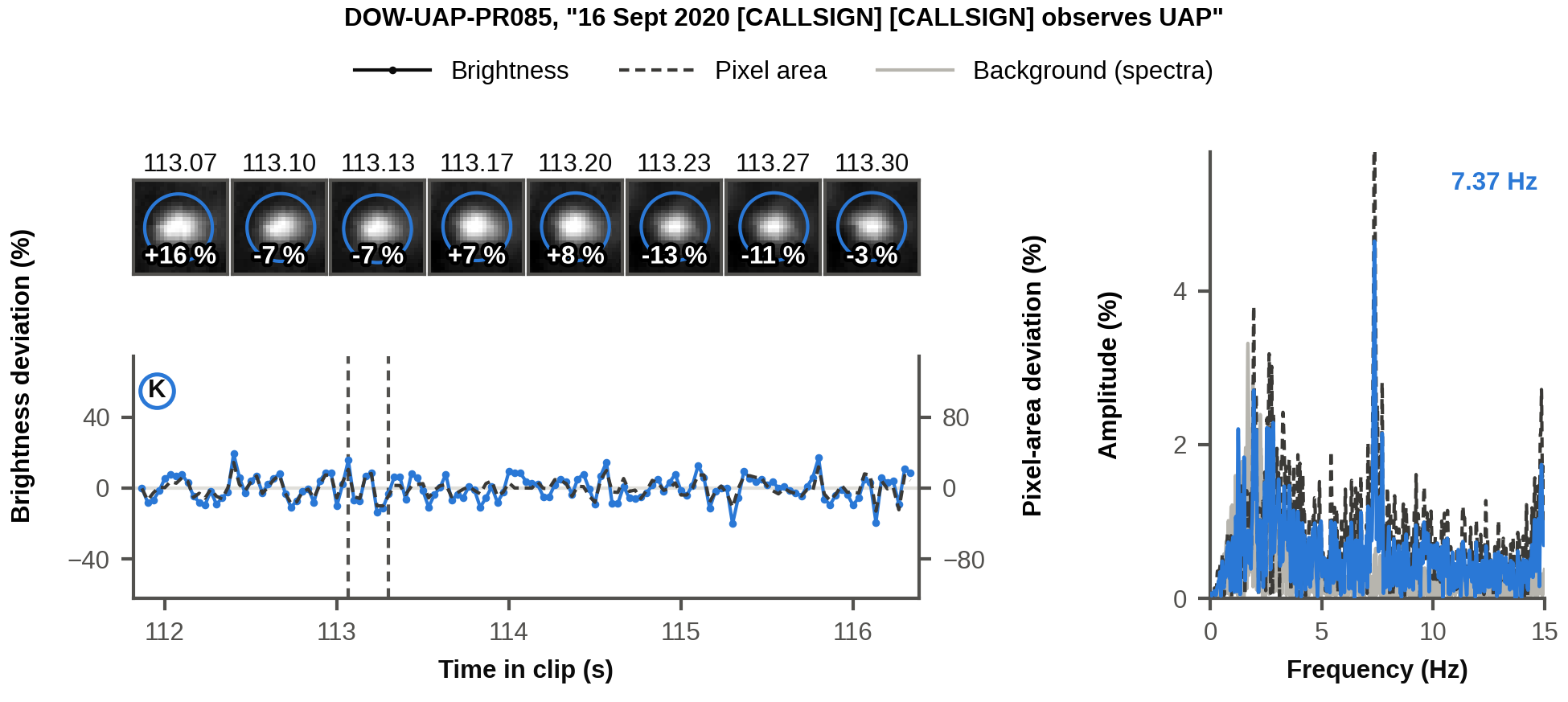}
\caption{\rev{Object K, a bright point in DOW-UAP-PR085, beats at 7.37~Hz. Video DOW-UAP-PR085, ``16 Sept 2020 [CALLSIGN] [CALLSIGN] observes UAP''~\cite{pr085video}.}}
\label{fig:pr085}
\end{figure}

\begin{figure}[!htbp]
\centering
\includegraphics[width=\textwidth]{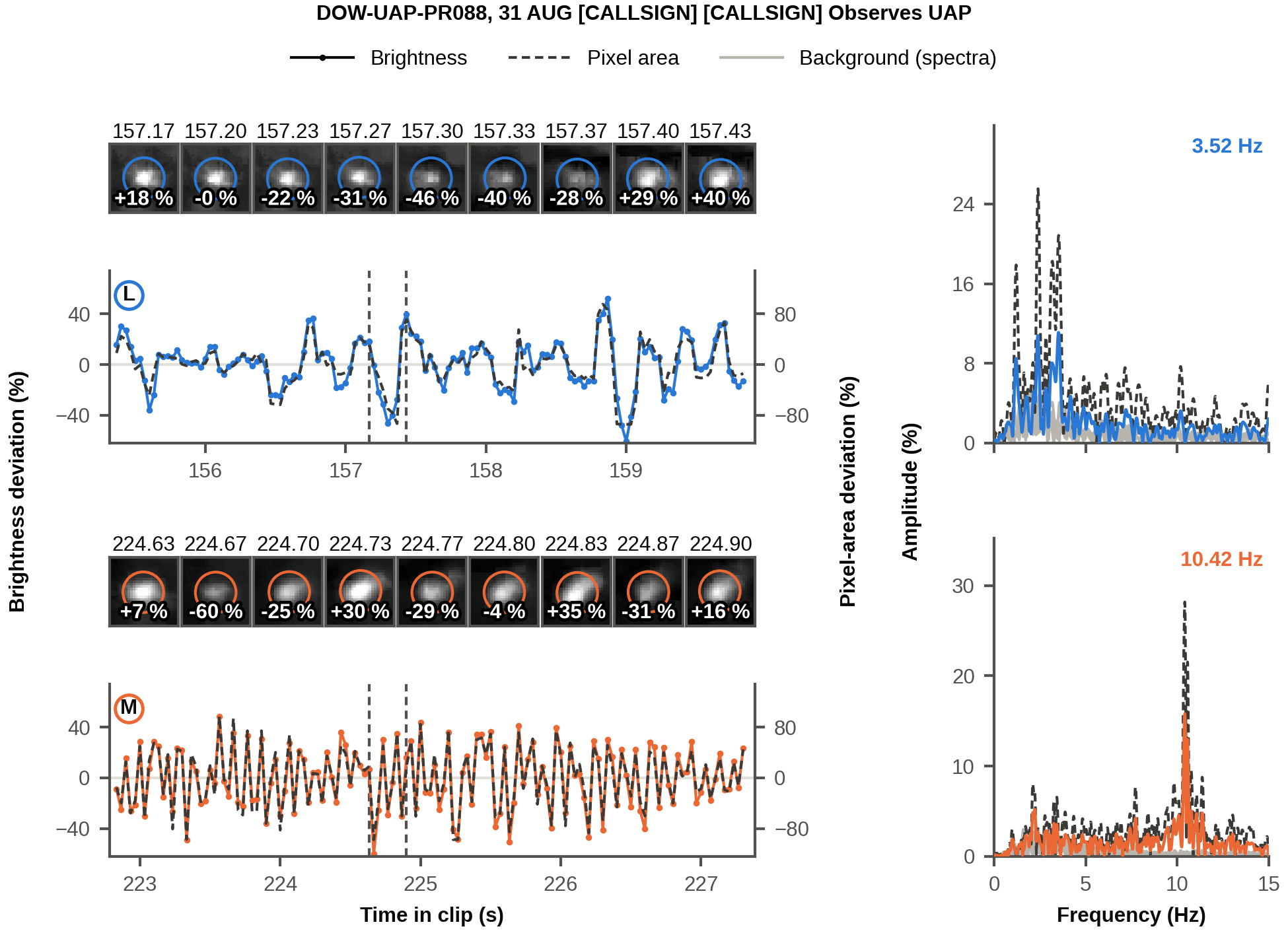}
\caption{\rev{Objects L and M, two bright points in DOW-UAP-PR088, beat at 3.5 and 10.4~Hz in the stretches shown. Video DOW-UAP-PR088, ``31 AUG [CALLSIGN] [CALLSIGN] Observes UAP''~\cite{pr088video}.}}
\label{fig:pr088}
\end{figure}

\begin{figure}[!htbp]
\centering
\includegraphics[width=\textwidth]{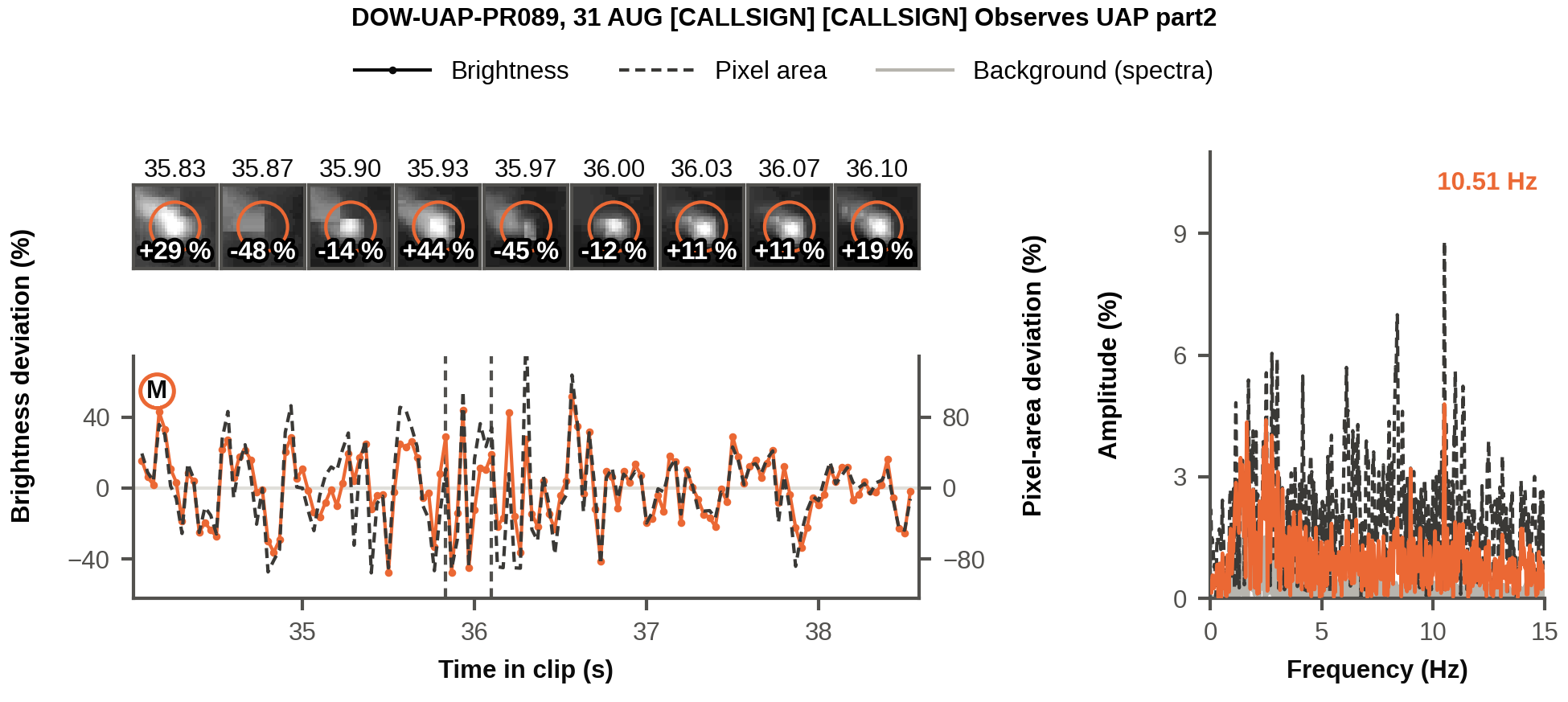}
\caption{\rev{Object M of DOW-UAP-PR088 (Supplementary Fig.~\ref{fig:pr088}) continues to beat at 10.5~Hz in DOW-UAP-PR089, the second part of the same recording. Video DOW-UAP-PR089, ``31 AUG [CALLSIGN] [CALLSIGN] Observes UAP part2''~\cite{pr089video}.}}
\label{fig:pr089}
\end{figure}

\begin{figure}[!htbp]
\centering
\includegraphics[width=\textwidth]{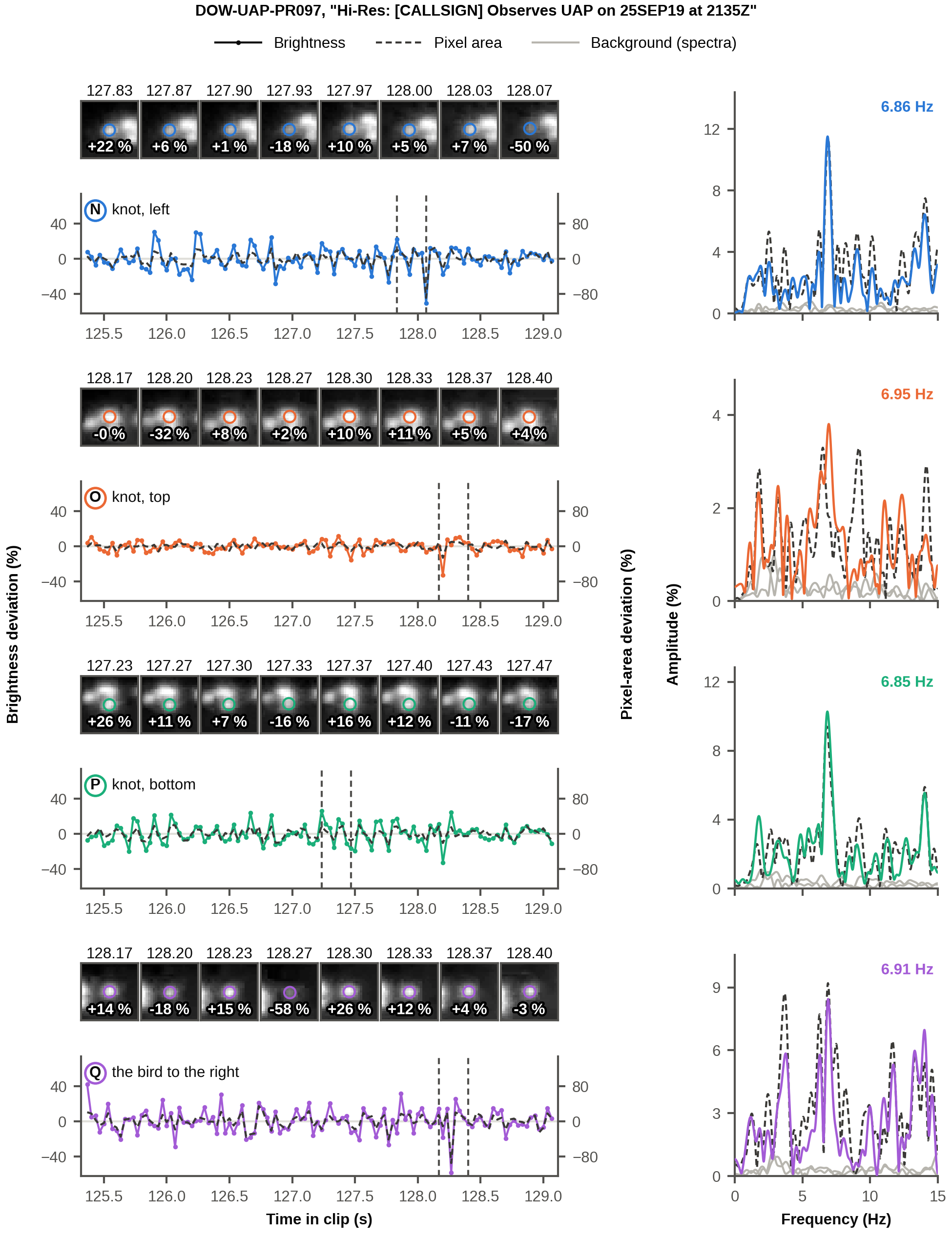}
\caption{\rev{Objects N--Q, four points flying together in DOW-UAP-PR097, beat at 6.85--6.95~Hz, each measured with the other three masked. Video DOW-UAP-PR097, ``Hi-Res: [CALLSIGN] Observes UAP on 25SEP19 at 2135Z''~\cite{pr097video}.}}
\label{fig:pr097}
\end{figure}

\begin{figure}[!htbp]
\centering
\includegraphics[width=\textwidth]{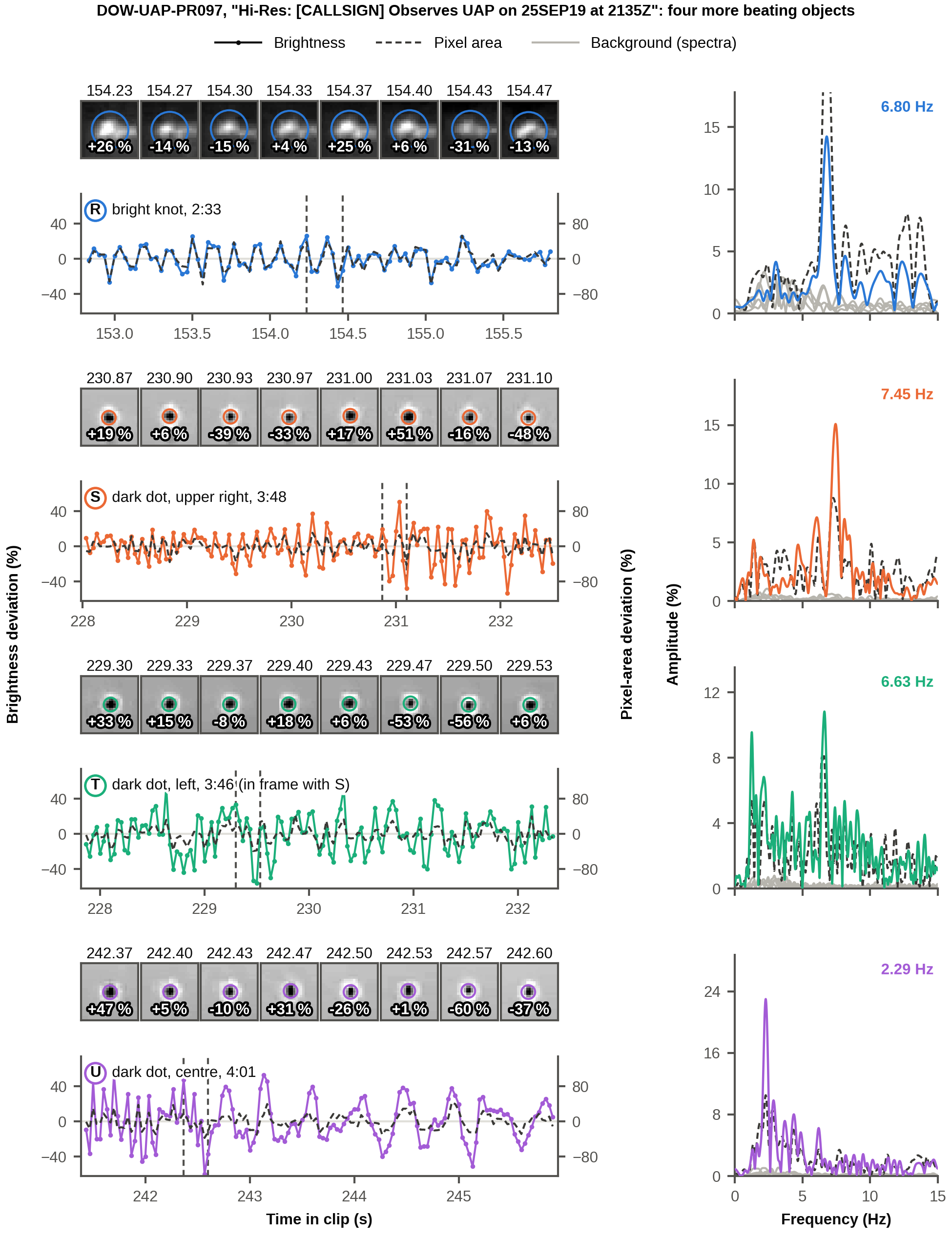}
\caption{\rev{Objects R--U, four more objects in DOW-UAP-PR097, beat at 2.29--7.45~Hz at other times in the clip. Video DOW-UAP-PR097, ``Hi-Res: [CALLSIGN] Observes UAP on 25SEP19 at 2135Z''~\cite{pr097video}.}}
\label{fig:pr097b}
\end{figure}

\begin{figure}[!htbp]
\centering
\includegraphics[width=\textwidth]{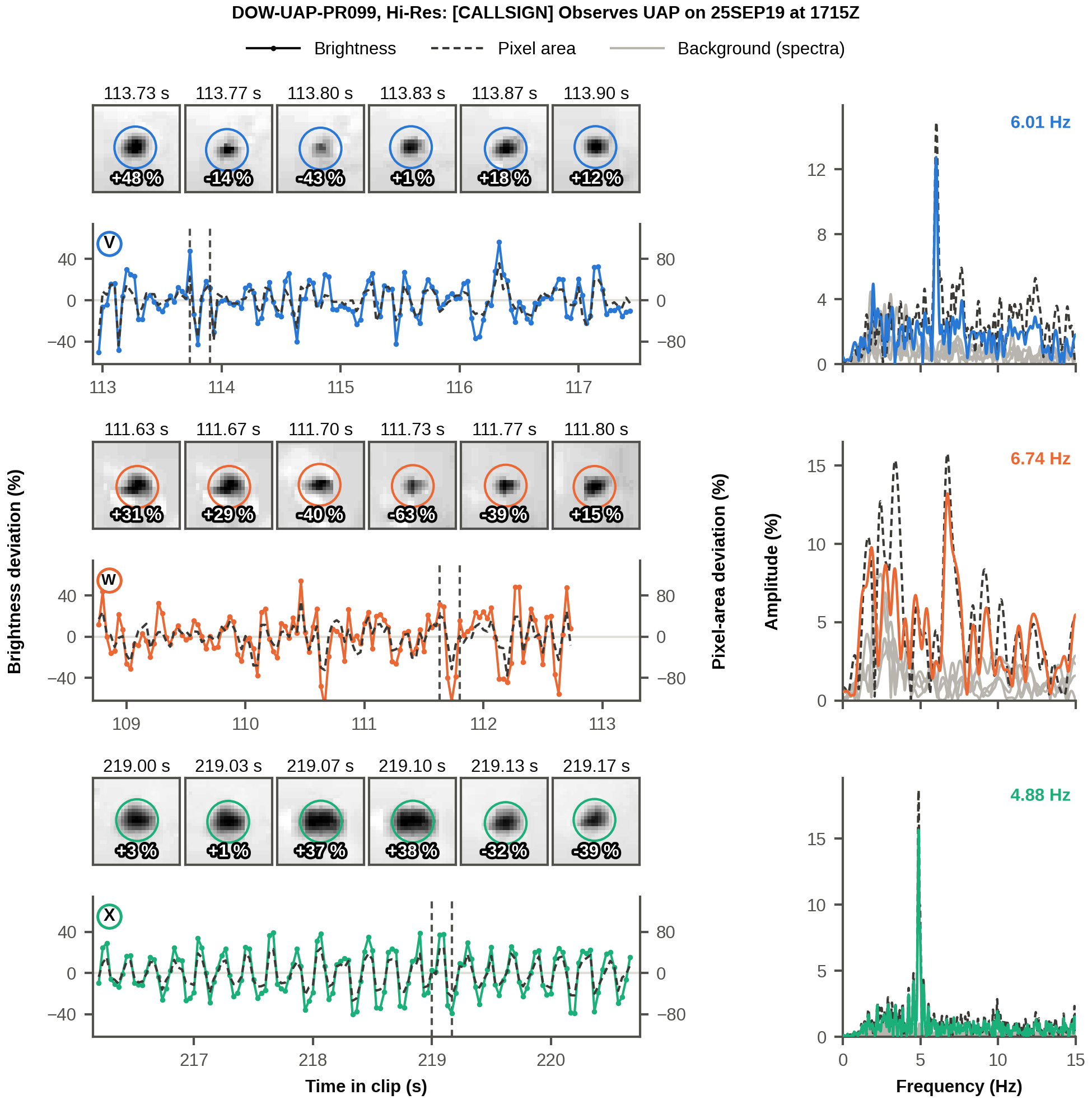}
\caption{\rev{Objects V--X, three dark points over the sea in DOW-UAP-PR099, beat at 4.9--6.7~Hz in the stretches shown. Video DOW-UAP-PR099, ``Hi-Res: [CALLSIGN] Observes UAP on 25SEP19 at 1715Z''~\cite{pr099video}.}}
\label{fig:pr099}
\end{figure}

\begin{figure}[!htbp]
\centering
\includegraphics[width=\textwidth]{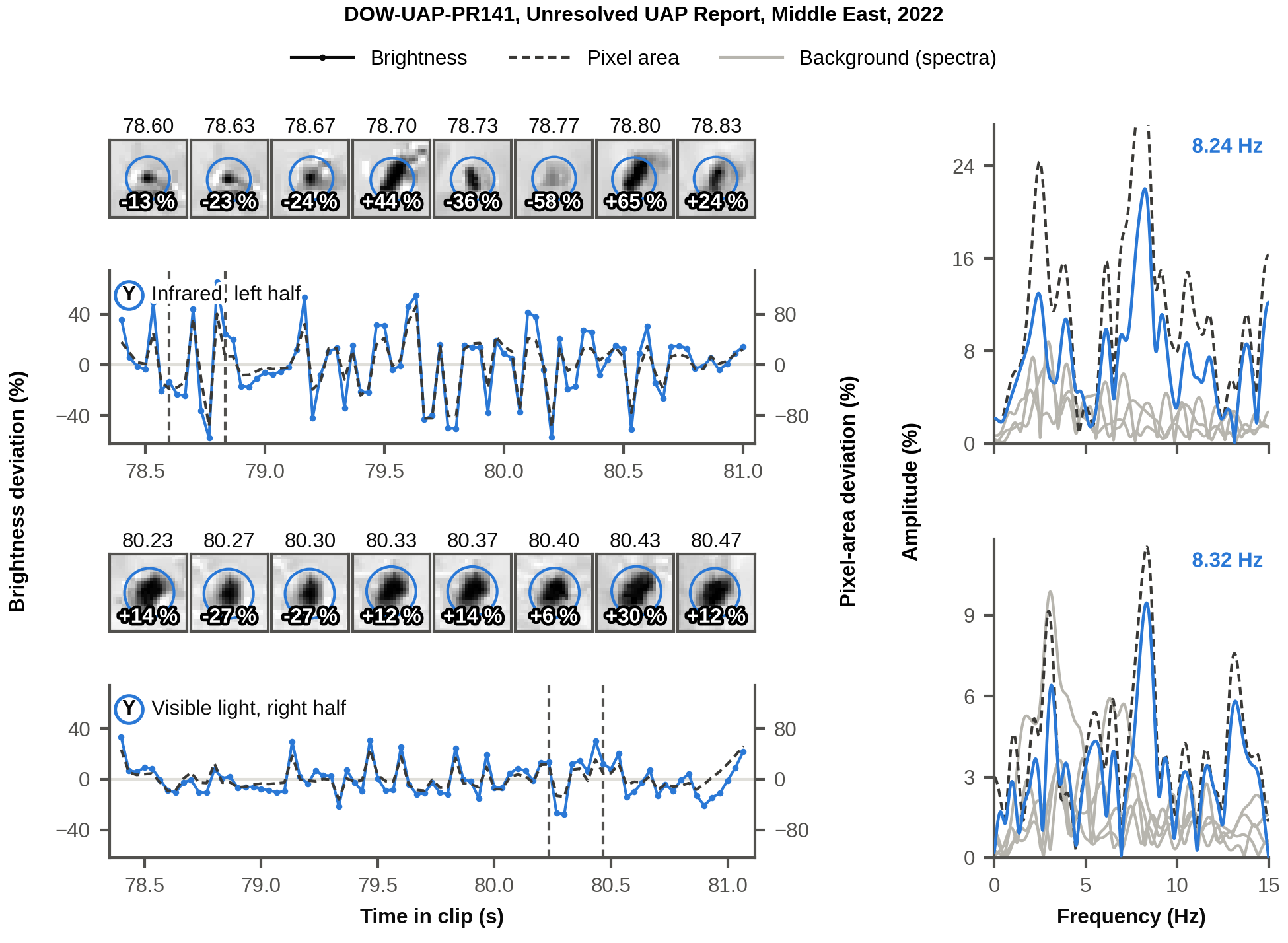}
\caption{\rev{Object Y, a dark point in DOW-UAP-PR141, beats at 8.2--8.3~Hz over the same frames in both halves of a split screen showing infrared (top) and visible light (bottom). Video DOW-UAP-PR141, ``Unresolved UAP Report, Middle East, 2022''~\cite{pr141video}.}}
\label{fig:pr141}
\end{figure}

\begin{figure}[!htbp]
\centering
\includegraphics[width=\textwidth]{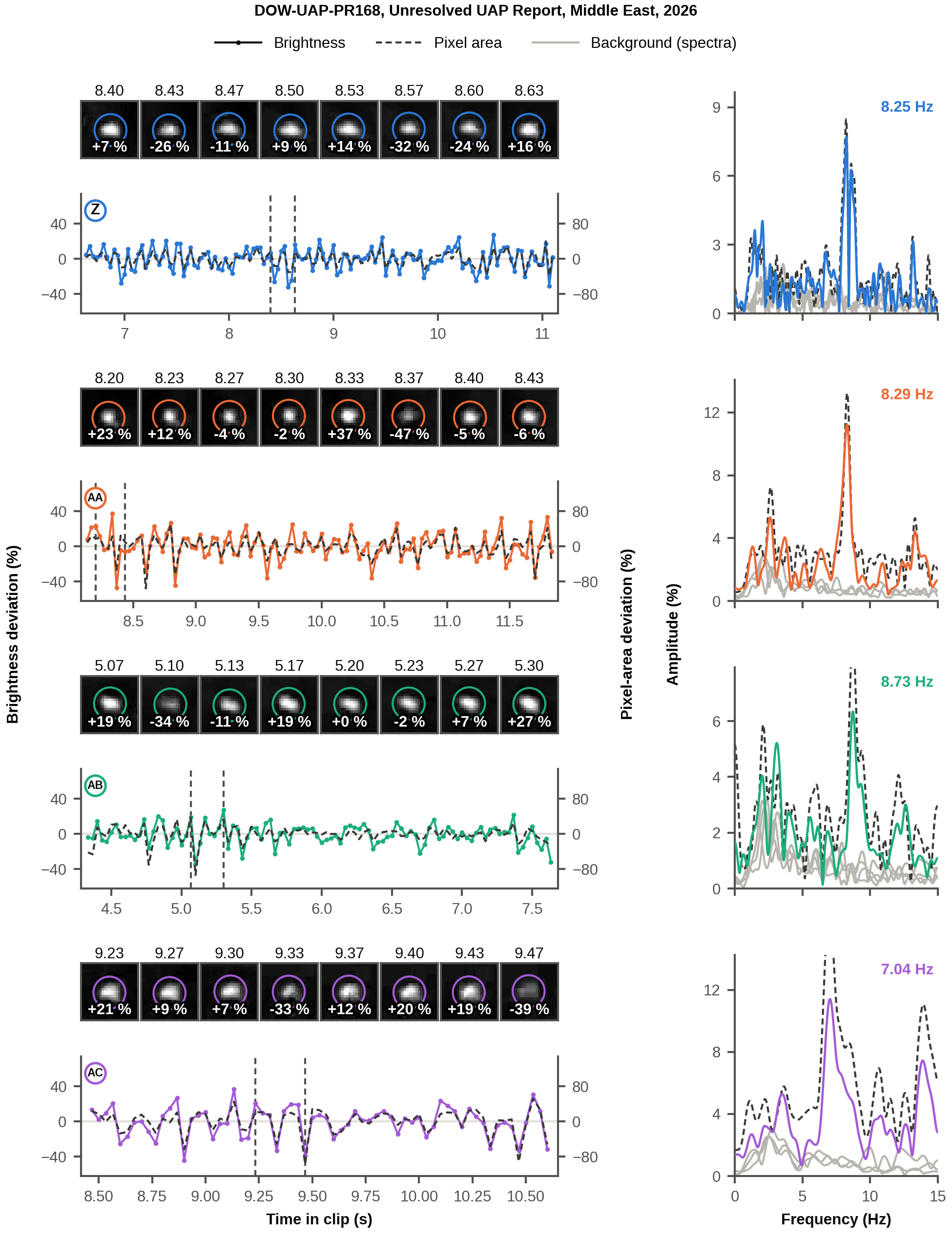}
\caption{\rev{Objects Z--AC, four points in view together in DOW-UAP-PR168, beat at 7.04--8.73~Hz, measured on the video's new pictures (the file shows every picture twice). Video DOW-UAP-PR168, ``Unresolved UAP Report, Middle East, 2026''~\cite{pr168video}.}}
\label{fig:pr168a}
\end{figure}

\begin{figure}[!htbp]
\centering
\includegraphics[width=\textwidth]{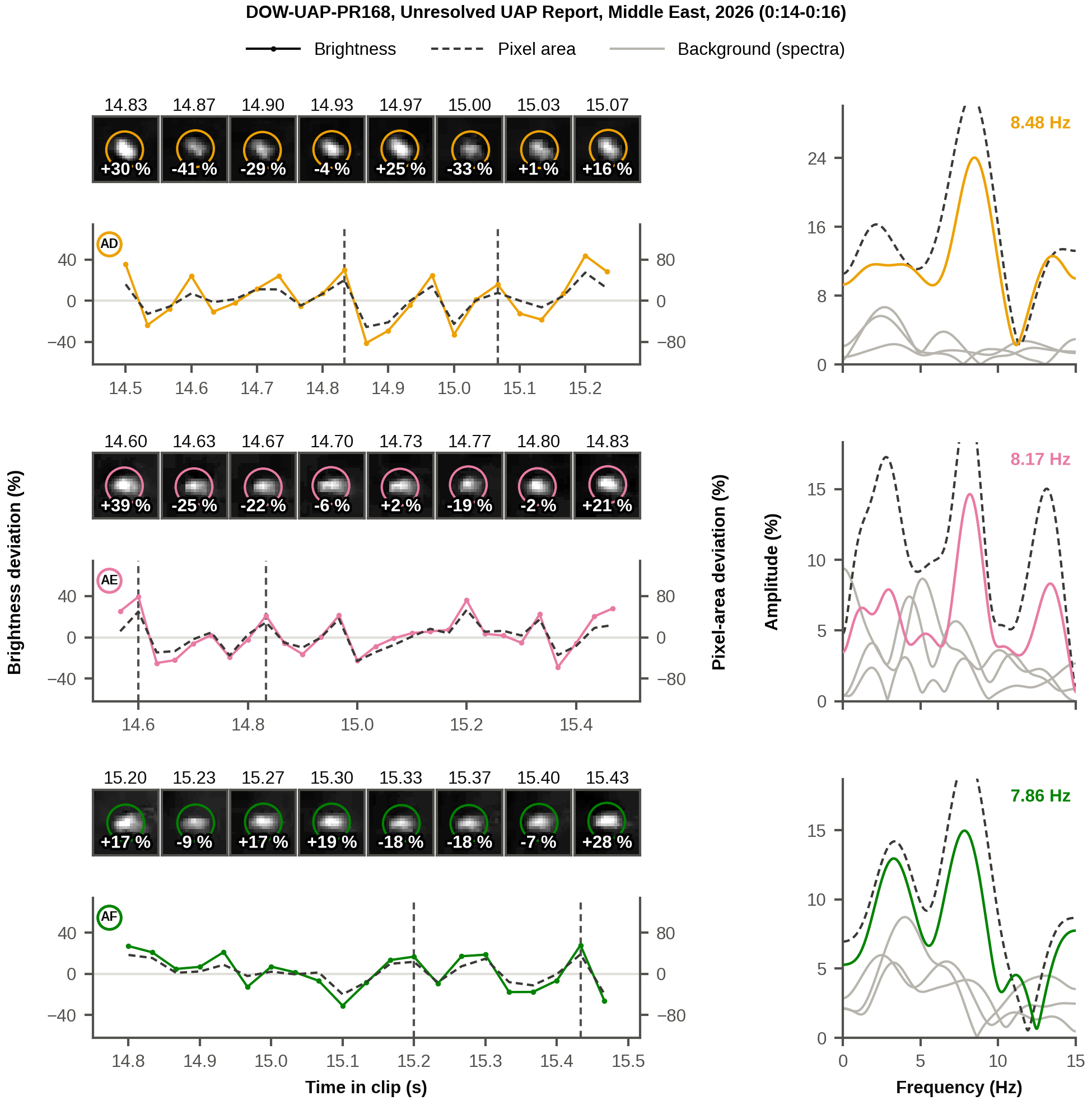}
\caption{\rev{Objects AD--AF, three more points in view together in DOW-UAP-PR168, beat at 7.86--8.48~Hz, each over about one second. Video DOW-UAP-PR168, ``Unresolved UAP Report, Middle East, 2026''~\cite{pr168video}.}}
\label{fig:pr168b}
\end{figure}

\begin{figure}[!htbp]
\centering
\includegraphics[width=\textwidth]{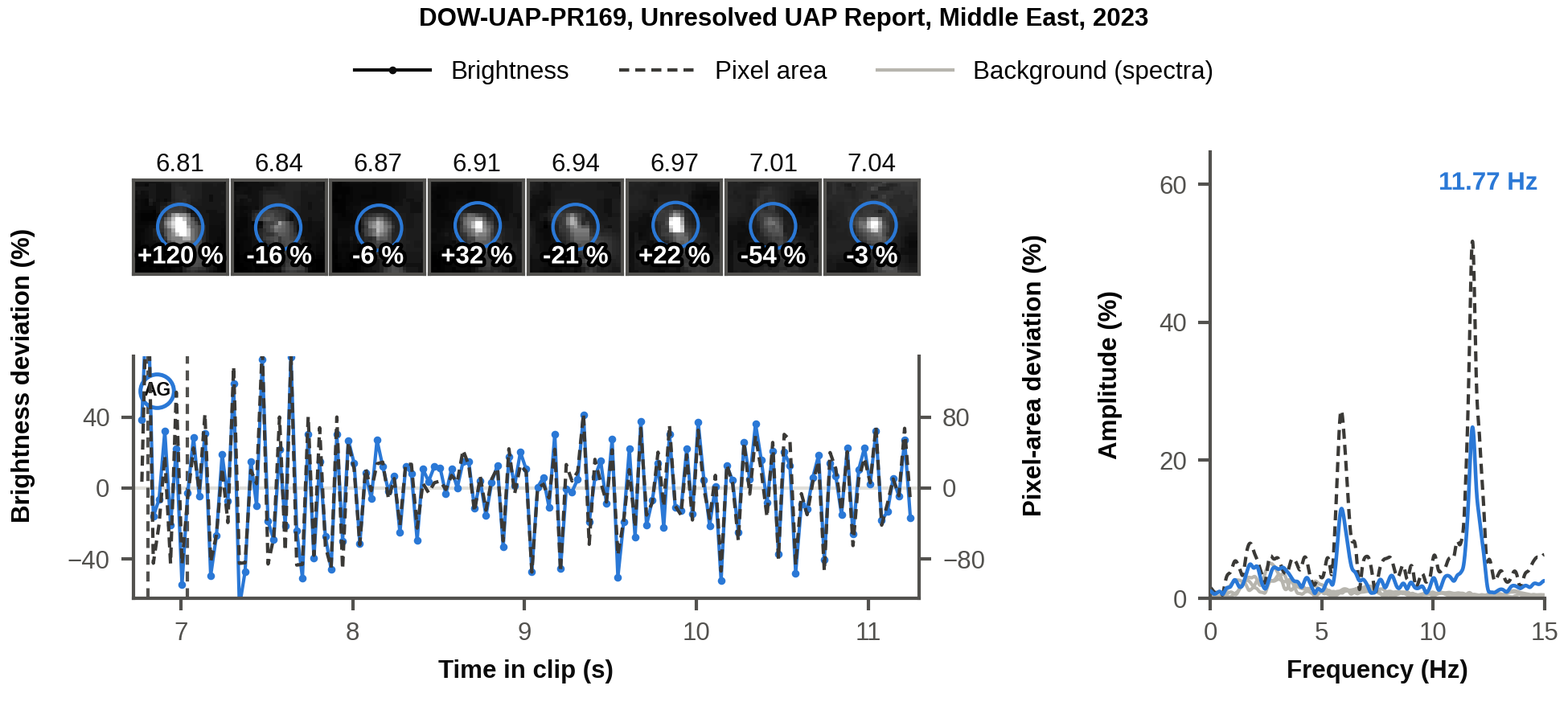}
\caption{\rev{Object AG, a point against a clear sky in DOW-UAP-PR169, beats most strongly at 11.77~Hz and also at half that rate, 5.9~Hz. Video DOW-UAP-PR169, ``Unresolved UAP Report, Middle East, 2023''~\cite{pr169video}.}}
\label{fig:pr169}
\end{figure}

\clearpage%
\bibliographystyle{unsrtnat}
\bibliography{refs,si_refs}